\documentclass[twocolumn,superscriptaddress,prb,showpacs]{revtex4-1}

\usepackage{amsmath,amssymb}
\usepackage{graphicx}
\usepackage{bm}
\usepackage[hypertex]{hyperref}
\usepackage{srcltx} 

\allowdisplaybreaks

\begin{document}

\title{
Time-reversal symmetric Kitaev model 
and topological superconductor in  
two dimensions
}
\date{\today}
\author{R. Nakai}
\affiliation{Department of Physics, University of Tokyo,
Tokyo 113-0033, Japan}
\email{rnakai@vortex.c.u-tokyo.ac.jp}
\author{S. Ryu}
\affiliation{Department of Physics, University of Illinois, 
1110 West Green St, Urbana IL 61801}
\author{A. Furusaki}
\affiliation{Condensed Matter Theory Laboratory, RIKEN, Wako,
Saitama 351-0198, Japan}

\begin{abstract}
A time-reversal invariant
Kitaev-type model
is introduced in which
spins (Dirac matrices) 
on the square lattice
interact via anisotropic nearest-neighbor
and next-nearest-neighbor exchange interactions.
The model is exactly solved by mapping it onto a tight-binding model
of free Majorana fermions coupled with static $\mathbb{Z}_2$
gauge fields.
The Majorana fermion model can be viewed as a model of
time-reversal invariant
superconductor and is classified as a member of symmetry class DIII
in the Altland-Zirnbauer classification.
The ground-state phase diagram has two topologically
distinct gapped phases which are distinguished
by a $\mathbb{Z}_2$ topological invariant.
The topologically nontrivial phase supports both a Kramers'
pair of gapless Majorana edge modes at the boundary and
a Kramers' pair of zero-energy Majorana states bound to
a $0$-flux vortex in the $\pi$-flux background.
Power-law decaying correlation functions of spins along the edge
are obtained by taking the gapless Majorana edge modes into account. 
The model is also defined on the one-dimension ladder, in which case again
the ground-state phase diagram has $\mathbb{Z}_2$ trivial and
non-trivial phases.
\end{abstract}


\maketitle

\section{Introduction}

Topological phases are a gapped state of matter which 
does not fall into a conventional characterization
of condensed matter systems in terms of symmetry breaking.
The prime and classic example is the fractional quantum Hall effect
which is realized in two-dimensional electron gas 
under strong magnetic field. 
Topological phases in the fractional quantum Hall effect
are characterized, e.g., 
by the presence of (chiral) edge states, 
by a set of fractionally charged quasiparticles
which obey fractional or non-Abelian statistics, 
and
also by the topological ground state degeneracy
when a system is put on a
spatial manifold with non-trivial topology.\cite{Wen}
A fractional quantum Hall state cannot be adiabatically deformed
into a trivial state of matter
such as an ordinary band insulator. 

While it is necessary to break time-reversal symmetry (TRS) 
to realize the fractional quantum Hall effect, 
a topological phase can exist without breaking TRS, 
as seen in several examples of gapped quantum spin liquid states
(e.g., $\mathbb{Z}_2$ spin liquid states). 
Furthermore,
a phase which is not topological, 
in the sense that it can be adiabatically connected to a trivial phase (vacuum),
can still be topologically distinct from the vacuum
once we impose some discrete symmetries,
such as TRS;
such phases can be called symmetry protected topological phase.
\cite{Chen, Gu2009, Pollmann2010}

Symmetry protected topological phases are recently
realized in the discovery of 
non-interacting topological band insulators, such as
the quantum spin Hall effect
and the three-dimensional topological insulator;\cite{HasanKane,QiZhang}
If we enforce TRS, these band insulators cannot be adiabatically connected
to a trivial band insulator, as seen from the presence of 
edge or surface states. 
Phases of non-interacting fermion systems (including 
Bogoliubov-de Genne quasiparticles in the presence of meanfield BCS
pairing gap)
have been fully classified in terms of presence or absence of 
discrete symmetries of various kind for arbitrary spatial
dimensions.\cite{Schnyder2008,Kitaev2009,Ryu2010}

Studies on realizations of strongly interacting counterparts
of these time-reversal symmetric topological band insulators, 
i.e., ``the fractional topological insulator,''
are still in their early stage.\cite{interactingTI} 

The notion of symmetry protected topological phases is not limited to
electron systems with TRS, but applies to bosonic systems
including quantum spin systems.\cite{Chen}
The Haldane phase in integer spin chains has been known as an example
of a gapped spin liquid phase in one spatial dimension 
with a localized end state which carries half-integer spin. 
It is recently uncovered that the Haldane phase 
has a symmetry protected topological order.\cite{Gu2009,Pollmann2010}

The list of experimentally established realizations of 
strongly interacting topological phases is still limited.
However, a number of exactly solvable models
have been proposed, 
helping us to deepen
our understanding of the topological orders in many-body systems. 
Examples are the Affleck-Kennedy-Lieb-Tasaki (AKLT) model,\cite{AKLT}
the quantum dimer models,\cite{RokhsarKivelson}
the toric code model,\cite{Kitaev2003}
and the string-net models,\cite{Levin2003} etc.
In Ref.\ \onlinecite{Kitaev2006}, 
Kitaev introduced an exactly solvable quantum spin model on
the two-dimensional honeycomb lattice.
A central feature of the honeycomb lattice Kitaev model,
among others, 
is that it realizes, in the absence of TRS, 
a gapped phase with 
a chiral Majorana edge state,
and non-Abelian anyonic excitations in the bulk.
Variants of the Kitaev model,
such as SU(2) invariant models,\cite{yaolee2011,Lai2011}
have been studied recently.\cite{Fiete2011}

In this paper, we consider an extension of the Kitaev
model on the square lattice that respects 
a TRS of some sort.
Following similar extensions of the Kitaev model on the
three-dimensional diamond lattice\cite{Ryu2009,CongjunWu08}
and on the two-dimensional square lattice,\cite{Yao08}
we consider 
two spin-1/2 degrees of freedom on each site that compose 
$4\times 4$ Dirac matrices ($\gamma$ matrices).
Similarly to the original Kitaev model, 
we consider interactions among spins 
which are anisotropic in space and are designed in such a way that 
the model is solvable through the (Majorana) fermion representation of spins. 
Written in terms of the fermions,
our model belongs to the symmetry class DIII in the Altland-Zirnbauer
classification of free fermions.
\cite{Schnyder2008,Altland1997}
Symmetry class DIII is a class of fermions which are 
subjected to TRS, and also to
particle-hole symmetry [i.e., a Majorana (or real) condition].
This should be contrasted with the original Kitaev model,
which when rewritten in terms of Majorana fermions, belongs to 
symmetry class D, which is a class of Majorana (real) fermions
without TRS.
One of our main findings is a topological phase which is characterized by 
the $\mathbb{Z}_2$ topological invariant of class DIII in the bulk,
and supports gapless {\it non-chiral} Majorana fermion edge modes
which form a Kramers pair:
the Bloch wavefunctions of the ``emergent'' Majorana fermions in this phase are
in the same topological class as those of 
fermionic quasiparticles in the topological superconductor
in symmetry class DIII.
This phase is a time-reversal symmetric analog of the non-Abelian phase
of the honeycomb lattice Kitaev model, and in fact, the model can be
viewed as a ``doubled'' version of the original Kitaev model;
just like the quantum spin Hall system with non-trivial $\mathbb{Z}_2$
topological invariant can be constructed 
from two copies of the integer quantum Hall systems with opposite chiralities.
From this point of view, our model is somewhat analogous 
to time-reversal invariant ``doubled'' anyon models discussed in 
Ref.\ \onlinecite{Gils2009}.

This paper is organized as follows.
In Sec.\ \ref{sec:model}, the Hamiltonian with nearest-neighbor and
next-nearest-neighbor interactions is presented in terms of
Dirac matrices and transformed to
free Majorana Hamiltonian that respects TRS. 
The symmetry class in the Altland-Zirnbauer classification is specified,
and the phase diagram of the ground states is obtained.
In Sec.~\ref{sec:2D_properties},
we show by numerical calculation and a topological argument
that the helical Majorana edge modes appear 
in the phase with a nontrivial $\mathbb{Z}_2$ topological invariant.
In Sec.~\ref{sec:spin_correlation}, some spin correlation functions
are calculated along the edge.
The existence of the gapless Majorana edge modes determines
the power-law decay of the correlation functions.
In Sec.~\ref{sec:vortex}, we confirm that an isolated vortex excitation of 
the $\mathbb{Z}_2$ gauge field hosts a time-reversal pair of zero-energy
Majorana bound states.
In Sec.~\ref{sec:1D_model}, we study the model on one-dimensional lattice.
Two distinct phases are found which are characterized by
the $\mathbb{Z}_2$ topological invariant.
In the Appendix we give an alternative representation of Dirac matrices
in terms of Jordan-Wigner fermions which keeps the same four-dimensional
Hilbert space at each site.

\section{Model}
\label{sec:model}

In this section, we introduce an extension of the Kitaev model that
respects time-reversal symmetry.  The Hamiltonian is written in terms
of Dirac matrices defined on each site of the two-dimensional square
lattice.  We first consider the Hamiltonian with nearest-neighbor
couplings only and show that its ground-state phase diagram has a
gapped phase and a gapless phase.\cite{Note_Yao}  We then add next-nearest-neighbor
couplings to the Hamiltonian, which open a gap in the gapless phase.
This gapped phase can be viewed as a topological superconducting phase
when the Hamiltonian is transformed
to a free Majorana
tight-binding Hamiltonian.
The time reversal symmetry is preserved in both of the gapped phases.

\subsection{Hamiltonian with nearest-neighbor interactions only}

There are a class of exactly solvable quantum spin models in which
Ising-type nearest-neighbor exchange interactions have different
easy-axis directions for each link on the lattice.  In the original
Kitaev model on the honeycomb lattice,\cite{Kitaev2006} three
components of the Pauli matrices are assigned to the three links
emanating from a site of the honeycomb lattice. Similarly, to define
an exactly solvable spin model on the square lattice, we can take
Dirac matrices and assign four components of the Dirac matrices to
four distinct types of links that emanate from each site, as in the
Kitaev-type model on the diamond lattice.\cite{Ryu2009}

\begin{figure}[t]
 \includegraphics[width=60mm]{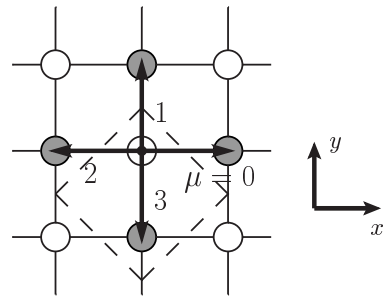}
 \caption{Square lattice and link vectors $\bm{e}_\mu$ with $\mu=0,1,2,3$
   emanating from a site on the A-sublattice (open circle) to a
   neighboring site on the B-sublattice (filled circle).  The dashed
   lines indicate a unit cell.}
 \label{square_lattice}
\end{figure}

The sites on the square lattice are divided into A- and B-sublattices.
Four links from a site on the A-sublattice are labeled, respectively, by
$\mu=0,1,2,3$ counterclockwise from the positive $x$-direction
(Fig.~\ref{square_lattice}).  Taking the lattice constant 
$a_0=1$, 
four
types of link vectors $\bm{e}_{\mu}$ are written in the two-dimensional
coordinate as
\begin{align}
 \bm{e}_0=
 \begin{pmatrix}
  1 \\ 0
 \end{pmatrix}
 ,\,\,
 \bm{e}_1=
 \begin{pmatrix}
  0 \\ 1
 \end{pmatrix}
 ,\,\,
 \bm{e}_2=
 \begin{pmatrix}
  -1 \\ 0
 \end{pmatrix}
 ,\,\,
 \bm{e}_3=
 \begin{pmatrix}
  0 \\ -1
 \end{pmatrix},
\end{align}
where the direction of each vector is chosen from a site of the
A-sublattice to a neighboring site of the B-sublattice.  In the
following the links labeled by $\mu(=0,1,2,3)$ are referred to as
``$\mu\text{-links}$''.

For each site on the square lattice,
we consider a four-dimensional bosonic Hilbert space. 
The four-dimensional Hilbert space can be considered as 
that of a spin-3/2 operator,\cite{Yao08} 
or the direct product of spin-1/2 degrees of freedom and two orbital degrees of freedom.
To describe the local bosonic Hilbert space, 
we define
a set of Dirac matrices $\alpha$
in terms of the $\gamma$ matrices
in the standard manner:\cite{Itzykson}
\begin{align}
 \alpha^0=\gamma^0, \quad
 \alpha^a=\gamma^0\gamma^a \quad (a=1,2,3).
\end{align}
The Dirac matrices $\alpha$ satisfy the anticommutation relations 
$\{\alpha^{\mu},\alpha^{\nu}\}=2\delta^{\mu\nu}$,
while the $\gamma$ matrices satisfy
$\{\gamma^{\mu},\gamma^{\nu}\}=2g^{\mu\nu}$,
where $g^{\mu\nu}=\text{diag}(1,-1,-1,-1)$.
With the fifth component of the $\gamma$ matrices,
$\gamma^5=i\gamma^0\gamma^1\gamma^2\gamma^3$,
we define another set of Dirac matrices $\zeta$,
\begin{align}
 \zeta^{0}=\gamma^5, \quad
 \zeta^a=\gamma^5\gamma^a
 \quad (a=1,2,3).
\end{align}
The $\zeta$ matrices also satisfy the anticommutation relations
$\{\zeta^{\mu},\zeta^{\nu}\}=2\delta^{\mu\nu}$.
We represent the $\gamma$ matrices as the direct product of
two Pauli matrices $\sigma^i$ and $\tau^i$
(the Dirac representation),
\begin{equation}
\gamma^0=\sigma^0\otimes\tau^3,
\quad
\gamma^a=i\sigma^a\otimes\tau^2
\quad (a=1,2,3),
\label{Dirac representation}
\end{equation}
where $\sigma^0$ and $\tau^0$ are $2\times2$ unit matrices.
The two sets of Dirac matrices are then written as
\begin{align}
 &\alpha^{\mu}:\,
  \alpha^0=\sigma^0\otimes\tau^3, \quad
  \alpha^a=\sigma^a\otimes\tau^1 \quad (a=1,2,3),
 \\
 &\zeta^{\mu}:\,
  \zeta^0=\sigma^0\otimes\tau^1, \quad
  \zeta^a=-\sigma^a\otimes\tau^3 \quad (a=1,2,3).
\end{align}

We introduce the nearest-neighbor spin Hamiltonian,
\begin{align}
\mathcal{H}_0=
-\sum_{\mu=0}^3J_{\mu}
 \sum_{\mu\text{-links}}
 (\alpha_j^{\mu}\alpha_k^{\mu}+\zeta_j^{\mu}\zeta_k^{\mu}) 
 \label{Hamiltonian-spin},
\end{align}
where $J_{\mu}$ is the coupling constant on $\mu$-links.  The
subscripts $j$ and $k$ refer to nearest-neighbor sites on the A- and
B-sublattice, respectively, which are connected by a
$\mu\text{-link}$.  
That is, the position vectors of the sites $j$ and
$k$, $\bm{r}_j$ and $\bm{r}_k$, are related by
$\bm{r}_k=\bm{r}_j+\bm{e}_\mu$.  Without loss of generality, we can
assume $J_\mu\ge0$.  
In terms of the two Pauli matrices $\sigma^{\mu}$
and $\tau^{\mu}$, the model can be written as 
\begin{align}
\mathcal{H}_0=
-\sum_{\mu=0}^3J_{\mu}
 \sum_{\mu\text{-links}}
 (\sigma_j^{\mu}\sigma_k^{\mu})
 (\tau_j^{3}\tau_k^{3}+\tau_j^{1}\tau_k^{1}). 
\end{align}
While the part of the exchange term involving the $\sigma$-matrices 
is anisotropic,
the part involving the $\tau$-matrices is isotropic and XY like;
the model has a U(1) symmetry rotating the $\tau$-matrices around 
the $\tau^2$ axis. 
This U(1) symmetry, however, will be lost when 
we later perturb 
the nearest neighbor model (\ref{Hamiltonian-spin}).

The model is also invariant under a kind of time-reversal symmetry 
operation
which is designed to become a time-reversal
symmetry operation for half-integer spin fermions in the Majorana
representation discussed later.
Let us first consider a time-reversal operation $T$ defined by
\begin{align}
& T= ({i}\sigma^2)\otimes({i}\tau^2) 
\mathcal{K},
\nonumber \\
&
T \sigma^a T^{-1} = -\sigma^a,
\quad
T \tau^a T^{-1} = -\tau^a, 
\end{align}
with complex conjugation operator $\mathcal{K}$ and
$a= 1,2,3$.
Note that $T^2 =+1$.
Under $T$, $\alpha$ and $\zeta$ are transformed as
\begin{align}
&
T \alpha^{\mu} T^{-1} = -\alpha_{\mu},
\quad
T \zeta^{\mu} T^{-1} = -\zeta_{\mu},
\quad
\nonumber \\
&
T {i}\gamma^5 \gamma^0 T^{-1}
=
-{i}\gamma^5 \gamma^0,
\end{align}
where covariant and contravariant vectors are
defined as
$\alpha^{\mu}=(\alpha^0,\alpha^a)$
and
$\alpha_{\mu}=(\alpha^0,-\alpha^a)$.
As we have noted,
while the $\sigma$-part of our Hamiltonian 
is fully anisotropic in $\sigma$ space,
the $\tau$-part of the Hamiltonian  
is invariant under a rotation around $\tau^2$ axis.
In particular, 
it is invariant under a rotation $R$
by $\pi/2$ around $\tau^2$ axis,
\begin{align}
R 
\left(
\begin{array}{c}
\tau^1 \\
\tau^2 \\
\tau^3 \\
\end{array}
\right)
R^{-1} = 
\left(
\begin{array}{c}
\tau^3 \\
\tau^2 \\
-\tau^1 \\
\end{array}
\right),
\quad
R=
\frac{
\tau^0
+
{i}\tau^2}
{\sqrt{2}}.
\end{align}
Under $R$, $\alpha$ and $\zeta$ are transformed as
\begin{align}
&
R\alpha^{\mu} R^{-1} = -\zeta^{\mu},
\quad
R\zeta^{\mu} R^{-1} = +\alpha^{\mu},
\nonumber \\
&
R {i}\gamma^5 \gamma^0
R^{-1} = +{i}\gamma^5 \gamma^0.
\end{align}
By combining $T$ with $R$
we can define yet another antiunitary operation, $T'=RT$,
\begin{align}
&
T' 
= RT
=
\frac{1}{\sqrt2}
({i}\tau^2 -\tau^0)
{i}\sigma^2
\mathcal{K},
\nonumber \\
&
T' \sigma^a T^{\prime -1} = -\sigma^a,
\quad
T'
\left(
\begin{array}{c}
\tau^1 \\
\tau^2 \\
\tau^3 \\
\end{array}
\right)
T^{\prime -1}
=
\left(
\begin{array}{c}
-\tau^3 \\
-\tau^2 \\
+\tau^1 \\
\end{array}
\right).
\end{align}
Below, 
with a slight abuse of language, 
we will call this operation $T'$
time-reversal operation. 
When applied to $\alpha$ and $\zeta$,
\begin{align}
&
T' \alpha^{\mu}T^{\prime -1}
=
+\zeta^{\ }_{\mu},
\quad
T'\zeta^{\mu}T^{\prime -1}
=
-\alpha^{\ }_{\mu},
\nonumber \\
&
T'
{i}\gamma^5 \gamma^0
T^{\prime -1}
=
-{i}\gamma^5 \gamma^0,
\label{eq: TRS on alpha and zeta}
\end{align}
i.e.,
time-reversal operation
$T'$ exchanges $\alpha$ and $\zeta$,
and covariant and contravariant vectors.
Notice that 
\begin{align}
T^{\prime 2} 
&=
{i}\tau^2,
\quad
T^{\prime 4} =
-1.
\end{align}
We will impose the time-reversal symmetry $T'$
throughout the paper.

The Hamiltonian (\ref{Hamiltonian-spin}) has the
integrals of motion defined for each plaquette $p$,
\begin{align}
W_p=\prod_{(j,k)\in p}\alpha_j^{\mu}\alpha_k^{\mu}
=\prod_{(j,k)\in p}\zeta_j^{\mu}\zeta_k^{\mu}, \label{plaquette_op}
\end{align}
where $(j,k)$ are the four links on the boundary of
a plaquette $p$, and
the sites $j$ and $k$ are on the A- and B-sublattices, respectively.

\subsection{Mapping to Majorana fermion model}

The honeycomb lattice Kitaev model can be mapped 
to a Majorana fermion problem in the presence of 
a $\mathbb{Z}_2$ gauge field
by representing the Pauli matrices in terms of four Majorana
fermions per site.\cite{Kitaev2006}
Similarly, we can represent the two sets of
Dirac matrices $\alpha^\mu$ and $\zeta^\mu$ with six Majorana fermions
$\lambda^p\,(p=0,\cdots,5)$:
\cite{Ryu2009,Levin2003,Yao08,CongjunWu08}
\begin{align}
 \alpha^{\mu}=i\lambda^{\mu}\lambda^4,\,\,\,
 \zeta^{\mu}=i\lambda^{\mu}\lambda^5, \label{gamma-Majorana}
\end{align}
where we have not written the site indices explicitly.
The Majorana fermions satisfy
$(\lambda^p)^\dagger=\lambda^p$ and
$\{\lambda^p,\lambda^{p'}\}=2\delta^{p p'}$.  
The bosonic Hamiltonian (\ref{Hamiltonian-spin}) 
is then mapped to, by using the relation (\ref{gamma-Majorana}),
a Majorana Hamiltonian 
\begin{align}
 \mathcal{H}_0=i\sum_{\mu=0}^3J_{\mu}\sum_{\mu\text{-links}}
 u_{jk}^{\mu}(\lambda_j^4\lambda_k^4+\lambda_j^5\lambda_k^5),
 \label
 {Hamiltonian-Majorana}
\end{align}
where $u_{jk}^{\mu}=i\lambda_j^{\mu}\lambda_k^{\mu}$ are defined on the
$\mu\text{-link}$ connecting two neighboring sites $j$ and $k$ which
belong to the A- and B-sublattices, respectively.  We will use
simplified notation $u_{jk}$ for $u_{jk}^{\mu}$ since $\mu$ is
uniquely determined by the neighboring sites $j$ and $k$.
The identity $(u_{jk})^2=1$ implies that the eigenvalue of $u_{jk}$
takes $\pm 1$.
The $u_{jk}$ defined on each link of the square lattice are
$\mathbb{Z}_2$ gauge fields.

Since $u_{jk}$ commute with each other and also with the Hamiltonian
(\ref{Hamiltonian-Majorana}), all $u_{jk}$ and the Hamiltonian can be
diagonalized simultaneously.  Hence the total Hilbert space 
$\mathcal{L}$
for
Majorana fermions is decomposed into subspaces 
$\mathcal{L}_{\{u_{jk}\}}$
which are specified by
the configurations of the eigenvalues of $\mathbb{Z}_2$ gauge fields
$u_{jk}$
on every link,
\begin{align}
 \mathcal{L}=\oplus \mathcal{L}_{\{u_{jk}\}}.
\end{align}
Within each subspace, the Hamiltonian is
regarded as a free Majorana fermion Hamiltonian, where $u_{jk}$ are
replaced by their eigenvalue $\pm 1$.

According to Lieb's theorem,\cite{Lieb1994} the energy of the free
Majorana Hamiltonian (\ref{Hamiltonian-Majorana}) is minimized when
$\mathbb{Z}_2$ gauge fields $u_{jk}$ are such that each plaquette has
a $\pi$-flux,
\begin{align}
 \prod_{(j,k)\in p}u_{jk}=-1. \label{Lieb_Th}
\end{align}
The left-hand side of Eq.\ (\ref{Lieb_Th}), which we denote
$\widetilde{W}_p $, is the Majorana 
fermion representation of the
plaquette operator $W_p$ in Eq.\ (\ref{plaquette_op}) and is
$\mathbb{Z}_2$ gauge invariant.
The condition (\ref{Lieb_Th}) is
satisfied, for example, by setting $u_{jk}=-1$ on the $0\text{-link}$s
and $u_{jk}=+1$ on the other links.
However, there is redundancy in the 
choice of $\mathbb{Z}_2$ gauge-field configuration for a given
flux configuration.

The time-reversal operation for Dirac matrices [Eq.\ (\ref{eq: TRS on alpha and zeta})] is translated
into that for Majorana fermions as
\begin{subequations}
\begin{align}
 T'
 \begin{pmatrix}
  \lambda^0 \\ \lambda^a
 \end{pmatrix}
 {T'}^{-1}=
 \begin{pmatrix}
  \lambda^0 \\ -\lambda^a
 \end{pmatrix},\,\,\,
 T'
 \begin{pmatrix}
  \lambda^4 \\ \lambda^5
 \end{pmatrix}
 {T'}^{-1}
 =
 \begin{pmatrix}
  -\lambda^5 \\ \lambda^4
 \end{pmatrix} \label{eq. tr for majorana1}
\end{align}
or
\begin{align}
 T'
 \begin{pmatrix}
  \lambda^0 \\ \lambda^a
 \end{pmatrix}
 {T'}^{-1}=
 \begin{pmatrix}
  -\lambda^0 \\ \lambda^a
 \end{pmatrix},\,\,\,
 T'
 \begin{pmatrix}
  \lambda^4 \\ \lambda^5
 \end{pmatrix}
 {T'}^{-1}
 =
 \begin{pmatrix}
  \lambda^5 \\ -\lambda^4
 \end{pmatrix}.
\label{eq. tr for majorana2}
\end{align}
\end{subequations}
In order to keep the $\mathbb{Z}_2$ gauge operators invariant under time-reversal
transformation, we  employ the two types of time-reversal rules to
Majorana fermions on each sublattice separately, i.e., Eq.\ (\ref{eq. tr for majorana1}) for
the A-sublattice and Eq.\ (\ref{eq. tr for majorana2}) for the B-sublattice.

\subsection{Projection}

The Majorana fermion representation (\ref{gamma-Majorana}) preserves the
commutation 
and
anticommutation relations of 
the
Dirac matrices $\alpha$ and $\zeta$.  
However, 
on each site,
the original four-dimensional Hilbert space is doubled in the Majorana 
fermion representation which employs
six flavors of
Majorana fermions (or, equivalently, three complex fermions),  
as in the original Kitaev model.
\cite{Kitaev2006} 
This redundancy can be removed by imposing the condition
at every site 
$l$ on the square lattice,
\begin{align}
 D_l:=i\prod_{p=0}^5\lambda^p_l=+1.
\label{D_l=1}
\end{align}
The operator $D_l$ is the Majorana fermion representation of 
$i\gamma^0_l\gamma^1_l\gamma^2_l\gamma^3_l\gamma^5_l$
that is a unit matrix by definition of the $\gamma$ matrices.
The condition (\ref{D_l=1}) is implemented by the projection operator
\begin{align}
 P=\prod_l\frac{1}{2}(1+D_l) \label{projection_operator}
\end{align}
acting on the states of 
the Majorana Hamiltonian.
(In the Appendix an alternative representation of Dirac matrices
is given in terms of Jordan-Wigner fermions which are free from
the redundancy.)

Now we show that the projection operator (\ref{projection_operator})
eliminates the arbitrariness of the choice of the $\mathbb{Z}_2$ gauge
field for a given flux configuration $\{\widetilde{W}_p\}$.
Let $|\Psi;\{u_{jk}\}\rangle$ be an eigenstate of Hamiltonian
(\ref{Hamiltonian-Majorana}) with a $\mathbb{Z}_2$ gauge-field
configuration $\{u_{jk}\}$.
It follows from the relation
\begin{align}
 [\mathcal{H}_0,D_l]=[\widetilde{W}_p,D_l]=0,
\end{align}
that $D_l|\Psi;\{u_{jk}\}\rangle$ is also an eigenstate of
$\mathcal{H}_0$ with the same flux configuration $\{\widetilde{W}_p\}$,
but with a different $\mathbb{Z}_2$ gauge-field configuration where the
$\mathbb{Z}_2$ gauge fields on the four links around the site $l$ are
multiplied by $-1$.  This can be seen from the relations
\begin{align}
 &\{u_{jk},D_j\}=\{u_{jk},D_k\}=0, \\
 &[u_{jk},D_l]=0 \quad (l\neq j,k). 
\end{align}
Furthermore, we can consider 
states
generated by acting $D_l$ on
multiple sites,
\begin{align}
 \prod_{l\in S}D_l|\Psi;\{u_{jk}\}\rangle, \label{D_state}
\end{align}
where $S$ is a set of sites from the square lattice.
One might think that
the total number of such states is $2^{N_\mathrm{tot}}$,
where $N_\mathrm{tot}$
is the total number of the lattice sites, since $(D_l)^2=1$.
However, under the periodic boundary condition,
the number of $\mathbb{Z}_2$ gauge-field configurations $\{u_{jk}\}$
generated in this way turns out to be
$2^{N_\mathrm{tot}-1}$, since the product of $D_l$ on the all sites,
\begin{align}
 \prod_l D_l\propto \prod_{(jk)}u_{jk}\prod_li\lambda_l^4\lambda_l^5,
 \label{D_all_sites}
\end{align}
does not change the $\mathbb{Z}_2$ gauge-field configurations.
Moreover,
eigenstates of the free Majorana fermion Hamiltonian are invariant
under the action of (\ref{D_all_sites}) up to an overall sign,
as creation/annihilation operators of single-particle states
anticommute with (\ref{D_all_sites}).
Obviously, the states generated by acting
$D_l$ from distinct sets $S$ and $S'$
are orthogonal,
\begin{align}
 \langle
 \Psi;\{u_{jk}\}
 |\prod_{l\in S}D_l \prod_{l'\in S'}D_{l'}
 |\Psi;\{u_{jk}\}
 \rangle
 =0,
\end{align}
unless $S=S'$ or $S'$ is the complementary set of $S$,
since the eigenvalues of the $\mathbb{Z}_2$ gauge-field
operators are different between two states.
Hence, the states of the form (\ref{D_state}) form 
$2^{N_\mathrm{tot}-1}$-dimensional orthonormal basis states.
Meanwhile, the number of flux configurations is $2^{N_\mathrm{tot}-1}$,
since the total flux must be unity ($\prod_p \widetilde{W}_p=1$).
Considering the fact that there are two additional, independent
integrals of motion defined on 
two closed loops $C_x,C_y$ going
around in the $x$- and $y$-directions,
\begin{subequations}
\begin{align}
 \widetilde{W}_{x}&=\prod_{(j,k)\in C_x}\alpha_j^{\mu}\alpha_k^{\mu}, \\
 \widetilde{W}_{y}&=\prod_{(j,k)\in C_y}\alpha_j^{\mu}\alpha_k^{\mu},
\end{align}
\end{subequations}
we find that
the number of $\mathbb{Z}_2$ gauge-field configurations for a given
local flux configuration ($\{\widetilde{W}_p\}$)
and global flux configurations ($\{\widetilde{W}_x,\widetilde{W}_y\}$),
is $2^{2N_\mathrm{tot}}/(2^{N_\mathrm{tot}-1}2^2)=2^{N_\mathrm{tot}-1}$.
Therefore the states (\ref{D_state}) exhaust the eigenstates
for all $\mathbb{Z}_2$ gauge-field configurations with the same flux
configuration.
Finally, projecting the state (\ref{D_state}) yields
\begin{align}
 P\prod_{l\in S}D_l|\Psi;\{u_{jk}\}\rangle=P|\Psi;\{u_{jk}\}\rangle,
 \label{projection_equiv}
\end{align}
since $(1+D_j)D_j=1+D_j$.
Equation (\ref{projection_equiv}) implies that the projected state
is independent of $\mathbb{Z}_2$ gauge choice.
Whatever $\mathbb{Z}_2$ gauge configuration is taken for 
a given flux configuration,
the same set of states are obtained after the projection;
any redundant state of the free Majorana fermion Hamiltonian
disappears after the projection.
Furthermore, we can conclude that matrix elements (for the projected
states) of gauge-invariant observables can be calculated by using
eigenstates of Majorana fermions with any particular $\mathbb{Z}_2$
gauge configuration.

\subsection{Phase diagram of the nearest-neighbor spin Hamiltonian}
\label{sec:nearest-neighbor}

Let us set $u_{jk}=-1$ on every $0\text{-link}$ and $u_{jk}=+1$ on the
other links of the square lattice, to satisfy the $\pi$-flux
condition, Eq.\ (\ref{Lieb_Th}).  This $\mathbb{Z}_2$ gauge-field
configuration, 
which we denote by a four-vector $u^\mu=(u^0,u^1,u^2,u^3)=(-1,1,1,1)$,
 preserves lattice translation
symmetry with the unit cell shown in Fig.~\ref{square_lattice}.
We introduce Fourier transformation of Majorana fermion operators
on the A-sublattice,
\begin{subequations}
\label{Fourier}
\begin{equation}
 a_{\bm{q}}^s=
 \frac{1}{\sqrt{2N}}\sum_{j\in\mathrm{A}} e^{-i\bm{q}\cdot\bm{r}_j}\lambda_j^s
 \qquad (s=4,5),
\end{equation}
and of those on the B-sublattice,
\begin{equation}
 b_{\bm{q}}^s=
 \frac{1}{\sqrt{2N}}\sum_{k\in\mathrm{B}} e^{-i\bm{q}\cdot\bm{r}_j}\lambda_k^s
 \qquad (s=4,5),
\label{Fourier-B}
\end{equation}
\end{subequations}
where $N$ is the number of unit cells,
and $\bm{r}_j$ in both of Eqs.\ (\ref{Fourier})
is the position vector of the site $j$ on the A-sublattice.
That is, $\bm{r}_j$ in Eq.\ (\ref{Fourier-B}) is related to the position
vector of the site $k$ on the B-sublattice by $\bm{r}_j=\bm{r}_k+\bm{e}_1$.
The inverse Fourier transform of Eqs.\ (\ref{Fourier}) is given by
\begin{subequations}
\begin{align}
\lambda_{j\in\mathrm{A}}^s
&
\equiv
a^s_{\boldsymbol{r}}
=\sqrt{\frac{2}{N}}\sum_{\bm{q}}
  e^{i\bm{q}\cdot\bm{r}_j} a_{\bm{q}}^s,
\\
\lambda_{k\in\mathrm{B}}^s
&
\equiv
b^s_{\boldsymbol{r}}
=\sqrt{\frac{2}{N}}\sum_{\bm{q}}
  e^{i\bm{q}\cdot(\bm{r}_k+\bm{e}_1)} b_{\bm{q}}^s,
\end{align}
\end{subequations}
where the wave vector $\bm{q}$ is in the first Brillouin zone,
$|q_x|+|q_y|\le\pi$.
The fermion operators defined in Eqs.\ (\ref{Fourier}) satisfy
the following relations:
\begin{subequations}
\begin{align}
&
a_{-\bm{q}}^s=\left(a_{\bm{q}}^s\right)^\dagger,
\qquad
b_{-\bm{q}}^s=\left(b_{\bm{q}}^s\right)^\dagger,
\\
&
\{a_{\bm{q}}^s,a_{\bm{q}'}^{s'}\}=
\{b_{\bm{q}}^s,b_{\bm{q}'}^{s'}\}=\delta_{\bm{q}+\bm{q}',0}\delta_{s,s'},
\quad
\{a_{\bm{q}}^s,b_{\bm{q}'}^{s'}\}=0.
\end{align}
\end{subequations}
One can thus regard $a_{\bm{q}}^s$ and $a_{-\bm{q}}^s$
($b_{\bm{q}}^s$ and $b_{-\bm{q}}^s$) as annihilation and
creation operators of fermions (or vice versa).
Hamiltonian (\ref{Hamiltonian-Majorana}) is written 
in the momentum space as
\begin{align}
\mathcal{H}_0 = &
\sum_{\bm{q}}
\left[
i\Phi(\bm{q})\left(
a^4_{-\bm{q}}b^4_{\bm{q}}+a^5_{-\bm{q}}b^5_{\bm{q}}
\right)
\right.\nonumber\\
&\left.{}\qquad
-i\Phi^*(\bm{q})\left(
b^4_{-\bm{q}}a^4_{\bm{q}}+b^5_{-\bm{q}}a^5_{\bm{q}}
\right)
\right],
\label{Hamiltonian-Majoranak}
\end{align}
where
\begin{align}
\Phi(\bm{q})&=
e^{iq_y}
\sum_{\mu}J_{\mu}u^{\mu}e^{i\bm{q}\cdot\bm{e}_{\mu}}
\nonumber\\
&=
e^{iq_y} (
-J_0e^{iq_x}+J_1e^{iq_y}+J_2e^{-iq_x}+J_3e^{-iq_y}
).
\label{Phi(q)}
\end{align}
The eigenvalues of (\ref{Hamiltonian-Majoranak}) are $E=\pm|\Phi(\bm{q})|$.
Each eigenstate is doubly degenerate, since $\lambda^4$ and $\lambda^5$ 
are decoupled in the Hamiltonian. 
The ground state is obtained by filling all the eigenstates with
negative energy.

\begin{figure}[t]
 \includegraphics[width=70mm]{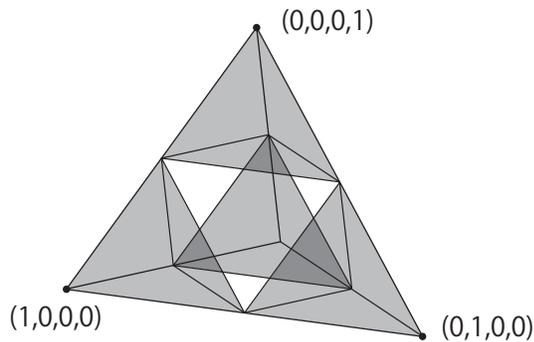}
 \caption{Phase diagram in the parameter space $(J_0,J_1,J_2,J_3)$.
Shaded regions are gapped phases and the other area is gapless phase.}
 \label{phase-diagram}
\end{figure}

The ground-state phase diagram is drawn in Fig.~\ref{phase-diagram},
where for illustration purpose we normalized the parameters
$\bm{J}=(J_0,J_1,J_2,J_3)$
such that
$J_0+J_1+J_2+J_3=1$, $J_\mu\ge0$.
The vertices of the (large) tetrahedron in Fig.~\ref{phase-diagram},
$\bm{J}=(1,0,0,0),(0,1,0,0),(0,0,1,0),(0,0,0,1)$,
correspond to the parameter sets in which
one of the four coupling constants is much stronger than the others.
On the edges of the tetrahedron the sum of two coupling constants
is equal to 1.
The four shaded regions (smaller tetrahedrons) in Fig.~\ref{phase-diagram}
are gapped phases in which there is an energy gap between positive
energy bands and negative energy bands.
The region including the isotropic point $\bm{J}=(1/4,1/4,1/4,1/4)$
(the non-shaded part in Fig.~\ref{phase-diagram}) is a gapless phase
where the positive and negative energy bands touch
at two Dirac points, around which
Majorana fermions have linear energy dispersions.
The gapless phase will become a gapped topological phase,
once an energy gap is opened by some perturbations,
as is the case in the Kitaev model.
On the boundary between a gapped phase and the gapless phase,
two Dirac points 
merge to become
a single point in the Brillouin zone.
This happens when one of the four $J_\mu$'s is equal to
the sum of the other three $J_\mu$'s.

\subsection{Hamiltonian with next-nearest-neighbor interaction}

We add, to the Hamiltonian $\mathcal{H}_0$, 
perturbations of the form of a product of Dirac matrices
from three neighboring sites.
As we will see, these perturbations will open a gap at the Dirac points
in the gapless phase.

\begin{figure}[t]
 \begin{center}
  \begin{tabular}[t]{cc}
   \begin{minipage}[c]{52mm}
    \includegraphics[width=52mm]{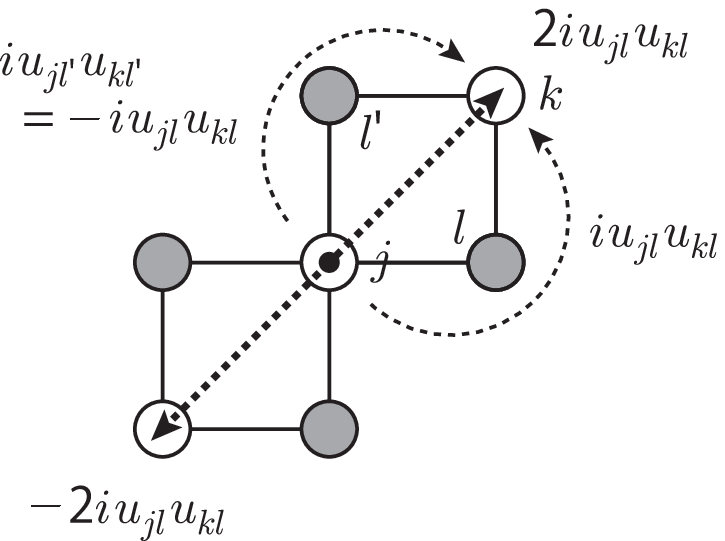}
   \end{minipage}&
   \begin{minipage}[c]{28mm}
    \includegraphics[width=28mm]{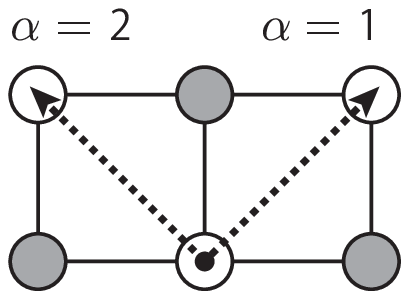}
   \end{minipage}\\
  \end{tabular}
 \end{center}
 \caption{Four-spin interaction terms. Left: The $\mathbb{Z}_2$ gauge
fields obtained from the two paths connecting next-nearest-neighbor
sites. Right: Two directions are labeled by $\alpha=1,2$,
respectively.}
 \label{nnn}
\end{figure}

Consider three neighboring sites $j$, $k$, and $l$ of
a single plaquette shown in Fig.~\ref{nnn}, where the sites
$j$ and $k$ belong to the same sublattice (either A or B).
We consider three-site interaction Hamiltonian of the form
\begin{equation}
\mathcal{H}_z=
\sum_{(jlk)}iK^{z}_{jlk}
(\alpha^\mu_j\alpha^\mu_l\alpha^\nu_l\alpha^\nu_k
-\zeta^\mu_j\zeta^\mu_l\zeta^\nu_l\zeta^\nu_k),
\label{three-sites-1}
\end{equation}
where the links $(jl)$ and $(lk)$ are a $\mu$-link
and a $\nu$-link, respectively.
In the Majorana 
fermion representation, the three-site interactions read as
\begin{subequations}
\label{product_of_4gammas_1}
\begin{align}
 & i (\alpha_j^{\mu}\alpha_l^{\mu})(\alpha_l^{\nu}\alpha_k^{\nu})
 = i u^{\mu}_{jl}u^{\nu}_{kl}\lambda_j^4\lambda_k^4, \\
 & i (\zeta_j^{\mu}\zeta_l^{\mu})(\zeta_l^{\nu}\zeta_k^{\nu})
 = i u^{\mu}_{jl}u^{\nu}_{kl}\lambda_j^5\lambda_k^5.
\end{align}
\end{subequations}
As the Majorana operators $\lambda_l^s$ do not appear explicitly
in the right hand side of Eqs.\ (\ref{product_of_4gammas_1}),
we can regard these perturbations as next-nearest-neighbor
hopping operators
for $\lambda^{4,5}_j$.
We have a different type of three-site interactions
in which different sets of Dirac matrices are used for two
links:
\begin{equation}
\mathcal{H}_x=
-\sum_{(jlk)}iK^{x}_{jlk}
(\alpha^\mu_j\alpha^\mu_l\gamma_l^5\gamma_l^0\alpha^\nu_l\alpha^\nu_k
-\zeta^\mu_j\zeta^\mu_l\gamma_l^5\gamma_l^0\zeta^\nu_l\zeta^\nu_k).
\label{three-sites-2}
\end{equation}
This yields another type of next-nearest-neighbor hopping term,
\begin{subequations}
\label{product_of_4gammas_2}
\begin{align}
 & i (\alpha_j^{\mu}\alpha_l^{\mu})
 \gamma^5_l\gamma^0_l(\zeta_l^{\nu}\zeta_k^{\nu})
 = - i u^{\mu}_{jl}u^{\nu}_{kl}\lambda_j^4\lambda_k^5, \\
 & i (\zeta_j^{\mu}\zeta_l^{\mu})
 \gamma^5_l\gamma^0_l(\alpha_l^{\nu}\alpha_k^{\nu})
 = i u^{\mu}_{jl}u^{\nu}_{kl}\lambda_j^5\lambda_k^4.
\end{align}
\end{subequations}
The summations in Eqs.\ (\ref{three-sites-1}) and
(\ref{three-sites-2}) are over any three neighboring sites which
belong to the same plaquette, including the three sites $(jl'k)$, in
addition to $(jlk)$, in Fig.~\ref{nnn}.
For the $\mathbb{Z}_2$ gauge fields satisfying the
$\pi$-flux condition, Eq.\ (\ref{Lieb_Th}), the product of
$\mathbb{Z}_2$ gauge fields $u_{jl}u_{kl}$ in
Eqs.\ (\ref{product_of_4gammas_1}) has the opposite sign compared to
the product $u_{jl'}u_{kl'}$.
We assume $K^{z}_{jlk}=-K^{z}_{jl'k}$
and $K^{x}_{jlk}=-K^{x}_{jl'k}$ so that the two paths give the
same contributions.
This leads to a vanishing next-nearest-neighbor
hopping for 0-flux plaquettes.

To summarize, under
the $\pi$-flux condition (\ref{Lieb_Th}),
we have the following two types of next-nearest-neighbor hopping
Hamiltonian:
\begin{subequations}
\begin{align}
 \mathcal{H}_z=&
-\sum_{j\in{\rm A}}
 \sum_{\alpha=1,2} i K^\alpha_z
 (\lambda_j^4 \lambda_{j+a_\alpha}^4 - \lambda_j^5 \lambda_{j+a_\alpha}^5)
 \nonumber\\
 &
+\sum_{j\in{\rm B}}
 \sum_{\alpha=1,2} i K^\alpha_z
 (\lambda_j^4 \lambda_{j+a_\alpha}^4 - \lambda_j^5 \lambda_{j+a_\alpha}^5),
\label{H_z}
 \\
 \mathcal{H}_x=&
-\sum_{j\in{\rm A}}
 \sum_{\alpha=1,2} i K^\alpha_x
 (\lambda_j^4 \lambda_{j+a_\alpha}^5 + \lambda_j^5 \lambda_{j+a_\alpha}^4)
 \nonumber\\
 &
+\sum_{j\in{\rm B}}
 \sum_{\alpha=1,2} i K^\alpha_x
 (\lambda_j^4 \lambda_{j+a_\alpha}^5 + \lambda_j^5 \lambda_{j+a_\alpha}^4),
\label{H_x}
\end{align}
\end{subequations}
where the parameters $K_{x,z}^{1,2}$ are hopping matrix elements.
The subscript $j+a_\alpha$ denotes the site located at
$\bm{r}_j+\bm{a}_\alpha$ with
$\bm{a}_1=(1,1)$ and $\bm{a}_2=(-1,1)$.
The two vectors $\bm{a}_{1,2}$ correspond to the two directions of
next-nearest-neighbor hopping
labelled by $\alpha=1,2$ in Fig.~\ref{nnn}.
We shall see that the next-nearest-neighbor hopping
terms of the form
\begin{align}
i(\lambda_j^4\lambda_k^4-\lambda_j^5\lambda_k^5)
\quad\mathrm{and}\quad
i(\lambda_j^4\lambda_k^5+\lambda_j^5\lambda_k^4)
\label{nnn-TR}
\end{align}
are invariant under time-reversal transformation
which is defined in the next section.

Fourier transformation of Eqs.\ (\ref{H_z}) and (\ref{H_x}) yields
\begin{subequations}
\begin{align}
\mathcal{H}_z=&
\sum_{\bm{q}}
\Theta_z(\bm{q})(
a^4_{-\bm{q}}a^4_{\bm{q}}-b^4_{-\bm{q}}b^4_{\bm{q}}
-a^5_{-\bm{q}}a^5_{\bm{q}}+b^5_{-\bm{q}}b^5_{\bm{q}}
),
\\
\mathcal{H}_x=&
\sum_{\bm{q}}
\Theta_x(\bm{q})(
a^4_{-\bm{q}}a^5_{\bm{q}}-b^4_{-\bm{q}}b^5_{\bm{q}}
+a^5_{-\bm{q}}a^4_{\bm{q}}-b^5_{-\bm{q}}b^4_{\bm{q}}
),
\label{Hamiltonian-Majoranannn}
\end{align}
\end{subequations}
where
\begin{equation}
\Theta_i(\bm{q})=K^1_i\sin(q_x+q_y)+K^2_i\sin(-q_x+q_y)
\end{equation}
for $i=z,x$.
The total Hamiltonian can be written as
\begin{equation}
\mathcal{H}=
\mathcal{H}_0+\mathcal{H}_z+\mathcal{H}_x
=\sum_{\bm{q}}\psi_{\bm{q}}^\dagger\chi(\bm{q})\psi_{\bm{q}},
\end{equation}
where
\begin{equation}
\psi_{\bm{q}}=
\begin{pmatrix}
a^4_{\bm{q}} \\ b^4_{\bm{q}} \\ a^5_{\bm{q}} \\ b^5_{\bm{q}}
\end{pmatrix}
\label{psi_q}
\end{equation}
and
\begin{align}
\chi(\bm{q})=&
\begin{pmatrix}
\Theta_z & i\Phi & \Theta_x & 0 \\
-i\Phi^* & -\Theta_z & 0 & -\Theta_x \\
\Theta_x & 0 & -\Theta_z & i\Phi \\
0 & -\Theta_x & -i\Phi^* & \Theta_z
\end{pmatrix}
\nonumber\\
 =& -\mathrm{Re}\,[\Phi(\bm{q})] \,c^y\otimes s^0
    -\mathrm{Im}\,[\Phi(\bm{q})]\,c^x\otimes s^0
\nonumber\\
 & +\Theta_z(\bm{q})\,c^z\otimes s^z+\Theta_x(\bm{q})\,c^z\otimes s^x.
 \label{Hamiltonian-Majorana-whole}
\end{align}
Here we have defined $c^{i}$ ($i=x,y,z$) as the Pauli matrices acting
on the sublattice indices
$(a,b)$, and $s^{i}$ as those on the Majorana flavors $(4,5)$.
The matrix $s^0$ is the $2\times2$ unit matrix
in the Majorana flavor space.
The Hamiltonian in the momentum space $\chi(\bm{q})$ is invariant
under the translation by reciprocal lattice vectors,
$\bm{G}=(\pm\pi,\pi)$. 
The eigenenergies are $\pm\varepsilon_{\bm{q}}$, where
\begin{equation}
\varepsilon_{\bm{q}}
=\sqrt{|\Phi(\bm{q})|^2+[\Theta_z(\bm{q})]^2+[\Theta_x(\bm{q})]^2}.
\label{varepsilon_q}
\end{equation}
Each energy level, for a given $\boldsymbol{q}$
has two-fold degeneracy.

We have seen in Sec.\ \ref{sec:nearest-neighbor} that the gapless phase
of Hamiltonian $\mathcal{H}_0$ has two Dirac points 
at
$\bm{q}=\bm{q}_i$ where
$\Phi(\bm{q}_i)=0$.
Non-vanishing matrix elements
$\Theta_x(\bm{q})$ and $\Theta_z(\bm{q})$
at the Dirac points give a band gap.
Hence the gapless phase is turned into a gapped phase by
including the three-spin interaction or next-nearest-neighbor hopping
interactions $\mathcal{H}_z$ and $\mathcal{H}_x$.

Incidentally, both $\Theta_x(\bm{q})$ and $\Theta_z(\bm{q})$ vanish
on the phase boundaries of the gapless and gapped phases of $\mathcal{H}_0$.
Thus, the phase boundaries do not change upon addition of
$\mathcal{H}_{z,x}$ to $\mathcal{H}_0$.

\subsection{Symmetries}
\label{subsec: Symmetries}

In this subsection, we consider symmetry properties of the Hamiltonian
$\mathcal{H}$ in the Majorana 
fermion
representation and show that it
belongs to DIII symmetry class of the Altland-Zirnbauer
classification.\cite{Altland1997}

To this end, we begin with transforming the Majorana Hamiltonian
$\mathcal{H}$ into Bogoliubov-de Gennes (BdG) Hamiltonian.
We define fermion creation and annihilation operators on site $j$
from the two flavors of Majorana fermions $\lambda^4_j$ and $\lambda^5_j$,
\begin{equation}
c_j=\frac{1}{2}(\lambda_j^4+i\lambda_j^5),
\qquad
c_j^\dagger=\frac{1}{2}(\lambda_j^4-i\lambda_j^5). \label{complex_fermion}
\end{equation}
Their Fourier transforms are written as
\begin{subequations}
\begin{equation}
\begin{pmatrix}
A_{\bm{q}} \\ A_{\bm{q}}^\dagger
\end{pmatrix}
=\frac{1}{\sqrt{N}}\sum_j
\begin{pmatrix}
e^{-i\bm{q}\cdot\bm{r}_j} c_j \\
e^{i\bm{q}\cdot\bm{r}_j} c_j^\dagger
\end{pmatrix},
\end{equation}
for the A sublattice 
($j\in \mathrm{A}$), and
\begin{equation}
\begin{pmatrix}
B_{\bm{q}} \\ B^\dagger_{\bm{q}}
\end{pmatrix}
=\frac{1}{\sqrt{N}}\sum_k
\begin{pmatrix}
e^{-i\bm{q}\cdot(\bm{r}_k+\bm{e}_1)} c_k \\
e^{i\bm{q}\cdot(\bm{r}_k+\bm{e}_1)} c_k^\dagger
\end{pmatrix},
\end{equation}
\end{subequations}
for the B sublattice 
($k\in \mathrm{B}$).
The Nambu field for the complex fermion is defined by
\begin{equation}
\Psi_{\bm{q}}=
\begin{pmatrix}
A_{\bm{q}} \\ B_{\bm{q}} \\ A^\dagger_{-\bm{q}} \\ B^\dagger_{-\bm{q}}
\end{pmatrix}
\end{equation}
and is related to $\psi_{\bm{q}}$ in Eq.\ (\ref{psi_q}) by
the unitary transformation,
\begin{equation}
\psi_{\bm{q}}=U\Psi_{\bm{q}},
\end{equation}
where the unitary matrix $U$ is given by
\begin{equation}
U=\frac{1}{\sqrt2}
\begin{pmatrix}
1 & 0 & 1 & ~0~ \\
0 & 1 & 0 & 1 \\
-i & 0 & i & 0 \\
0 & -i & 0 & i
\end{pmatrix}.
\end{equation}
Then the Hamiltonian $\mathcal{H}$ can be written 
in the form of BdG Hamiltonian,
\begin{equation}
\mathcal{H}=
\sum_{\bm{q}}\Psi_{\bm{q}}^\dagger\tilde{\chi}(\bm{q})\Psi_{\bm{q}},
\end{equation}
where
\begin{align}
\tilde{\chi}(\bm{q})=&U^\dagger\chi(\bm{q})U
\nonumber\\
=&
\begin{pmatrix}
0 & i\Phi(\bm{q}) & \Theta(\bm{q}) & 0 \\
-i\Phi^*(\bm{q}) & 0 & 0 & -\Theta(\bm{q}) \\
\Theta^*(\bm{q}) & 0 & 0 & i\Phi(\bm{q}) \\
0 & -\Theta^*(\bm{q}) & -i\Phi^*(\bm{q}) & 0
\end{pmatrix} \label{chi_tilde}
\\
=&
-\mathrm{Re}\,[\Phi(\bm{q})] c^y \otimes t^0
-\mathrm{Im}\,[\Phi(\bm{q})] c^x \otimes t^0
\nonumber\\
&
+\mathrm{Re}\,[\Theta(\bm{q})] c^z \otimes t^x
-\mathrm{Im}\,[\Theta(\bm{q})] c^z \otimes t^y
\end{align}
with $\Theta(\bm{q})$ defined by
\begin{equation}
\Theta(\bm{q})=\Theta_z(\bm{q})+i\Theta_x(\bm{q}).
\end{equation}
The Pauli matrices $t^i$ ($i=x,y,z$) and the $2\times2$ unit matrix $t^0$
act on the Nambu indices. 

We are ready to discuss symmetries of our model in terms of the BdG
Hamiltonian $\tilde\chi(\bm{q})$.
Under the particle-hole transformation generated by
$\mathcal{P}=t^x\mathcal{K}$,
where $\mathcal{K}$ is complex conjugation operator,
the BdG Hamiltonian changes its sign,
\begin{equation}
t^x \tilde{\chi}^T(-\bm{q}) t^x = -\tilde{\chi}(\bm{q}).
\label{PHS_BdG}
\end{equation}
We note that $\mathcal{P}^2=+1$.

The BdG Hamiltonian is invariant under time-reversal transformation
\begin{equation}
i(c^z \otimes t^y)\tilde{\chi}^T(-\bm{q})(-i)(c^z \otimes t^y)
=\tilde{\chi}(\bm{q}).
\label{TRS_BdG}
\end{equation}
The time-reversal operator $\mathcal{T}=c^z\otimes it^y \mathcal{K}$ 
obeys $\mathcal{T}^2=-1$.

We conclude from these symmetry properties that the BdG Hamiltonian
$\tilde{\chi}$ belongs to symmetry class DIII;
see, e.g., Table 1 in Ref.~\onlinecite{Ryu2010}.
It is known from the classification theory of topological insulators
and superconductors\cite{Schnyder2008,Ryu2010,Kitaev2009} that
gapped ground states of class DIII Hamiltonian in two spatial dimensions
can be classified by a $\mathbb{Z}_2$ index;
see, e.g., Table 3 in Ref.~\onlinecite{Ryu2010}.

The product of particle-hole and time-reversal transformations,
$\mathcal{TP}$, yields
\begin{equation}
(c^z \otimes t^z)\tilde{\chi}(\bm{q})(c^z \otimes t^z)
=-\tilde{\chi}(\bm{q}),
\end{equation}
i.e., $\tilde{\chi}(\bm{q})$ anticommutes with $c^z\otimes t^z$.
In the basis where $c^z\otimes t^z$ is
$\mathrm{diag}(1,1,-1,-1)$, the BdG Hamiltonian is written
in the off-diagonal form,
\begin{equation}
\tilde{\chi}_D(\bm{q})=I_{24}\tilde{\chi}(\bm{q})I_{24}=
\begin{pmatrix}
0 & D(\bm{q}) \\ D^\dagger(\bm{q}) & 0
\end{pmatrix},
\label{chiral_BdG}
\end{equation}
where
\begin{equation}
I_{24}=
\begin{pmatrix}
1 & 0 & 0 & 0 \\
0 & 0 & 0 & 1 \\
0 & 0 & 1 & 0 \\
0 & 1 & 0 & 0
\end{pmatrix}
\end{equation}
and
\begin{equation}
D(\bm{q})=
\begin{pmatrix}
\Theta(\bm{q}) & i\Phi(\bm{q}) \\
-i\Phi^*(\bm{q}) & -\Theta^*(\bm{q})
\end{pmatrix}.
\label{D(q)}
\end{equation}
Since $I_{24}$ and $t^x$ commute, we find from Eqs.\ (\ref{PHS_BdG})
and (\ref{chiral_BdG}) that $D(\bm{q})$ satisfies the skew relation
\begin{equation}
D^T(-\bm{q})=-D(\bm{q}).
\label{skew}
\end{equation}
Since $I_{24}c^z\otimes it^y I_{24}=it^y$, the time-reversal operator
for $\tilde{\chi}_D(\bm{q})$ is $\mathcal{T}=it^y\mathcal{K}$.

The symmetry relations in Eqs.\ (\ref{PHS_BdG}) and (\ref{TRS_BdG})
lead to symmetry relations for the Majorana Hamiltonian
$\chi(\bm{q})$ through the unitary transformation.
The particle-hole symmetry relation implies
\begin{equation}
\chi^T(-\bm{q})=-\chi(\bm{q}),
\label{PHS_Majorana}
\end{equation}
while the time-reversal symmetry gives
\begin{equation}
c^z\otimes(is^y)\chi^T(-\bm{q})c^z\otimes(-is^y)=\chi(\bm{q}).
\label{TRS_Majorana}
\end{equation}
It follows from these relations that $\chi(\bm{q})$ anticommutes
with $c^z\otimes s^y$,
\begin{equation}
(c^z\otimes s^y)\chi(\bm{q})(c^z\otimes s^y)=-\chi(\bm{q}).
\label{chiral_Majorana}
\end{equation}
We note that the time-reversal operator $c^z\otimes(is^y)\mathcal{K}$
in Eq.\ (\ref{TRS_Majorana}) is consistent with that
for the Majorana operators [Eqs.\ (\ref{eq. tr for
majorana1}) and (\ref{eq. tr for majorana2})].

We return to the time-reversal symmetry of the next-nearest-neighbor
hopping terms, Eq.\ (\ref{nnn-TR}).
When the time-reversal operator 
$\mathcal{T}=c^z\otimes(is^y)\mathcal{K}$
is applied to
the $\lambda_j^s\lambda_k^{s'}$,
$c^z$ does nothing since both sites $j$ and $k$ are on the same sublattice,
while $is^y$ interchanges $s=4$ and $s=5$ with a factor of $-1$ ($+1$)
for $s\neq s'$ ($s=s'$).
Thus,
\begin{subequations}
\begin{align}
 &\mathcal{T}i\lambda_j^4\lambda_k^5\mathcal{T}^{-1}=i\lambda_j^5\lambda_k^4, \\
 &\mathcal{T}i\lambda_j^4\lambda_k^4\mathcal{T}^{-1}=-i\lambda_j^5\lambda_k^5,
\end{align}
\end{subequations}
and the hopping terms in Eq.\ (\ref{nnn-TR}) are invariant
under the time-reversal transformation.

\section{Phases and topological invariant}
\label{sec:2D_properties}

In this section we show the existence of a Kramers' pair of Majorana edge
modes in the 
topological phase and define a $\mathbb{Z}_2$ index
that distinguishes between the topologically nontrivial and trivial
phases.

\subsection{Energy spectrum}
\label{sec:energy spectrum}

We examine the energy spectrum of Majorana fermions by varying
two coupling constants $J_0$ and $J_1$
while the others are kept fixed as $J_2=J_3=J$
and $K_x^1=K_x^2=K_z^1=K_z^2=K$, for simplicity.
All coupling constants are taken to be positive.
We set the $\mathbb{Z}_2$ gauge fields $u^{\mu}=(-1,1,1,1)$ as before.

\begin{figure}[t]
 \includegraphics[width=80mm]{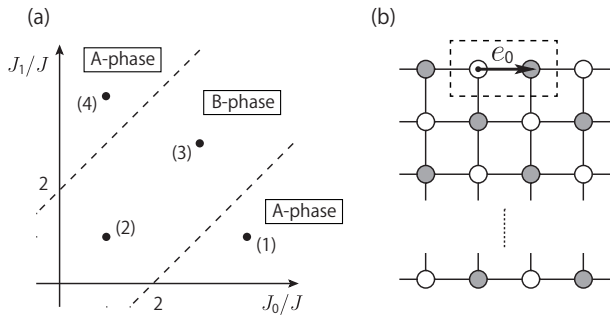}
\caption{(a) Phase diagram in the $J_0/J-J_1/J$ plane.
The region between the dashed lines $J_1=J_0\pm2J$, including
the isotropic point $J_1=J_0=J$, is the gapless phase when
the next-nearest-neighbor hopping terms are absent
(``B-phase'').
The gapless phase is turned into a topologically
nontrivial phase when the second-nearest-neighbor hopping is turned on.
Both the gapped phases for $J_1>J_0+2J$ and for $J_1<J_0-2J$
remain topologically trivial upon including the second-nearest-neighbor
hopping terms
(``A-phase'').
The numbered solid circles indicate the parameter sets for
which the energy spectra are calculated and shown
in Fig.~\ref{energy_spectrum-2Dstrip}.
(b) Strip geometry with edges along the vector $\bm{e}_0$.
}
\label{strip_phase}
\end{figure}

When $K=0$, the eigenvalues of $\chi(\bm{q})$ are given by
\begin{align}
E_{\bm{q}}=\pm|\Phi(\bm{q})|
 =\pm\left|\sum_{\mu=0}^3J_{\mu}u^{\mu}
  e^{i\bm{q}\cdot\bm{e}_{\mu}}\right|.
\end{align}  
When 
$J_0-2J<J_1<J_0+2J$ 
the positive and negative energy bands touch
at two Dirac points
(``B-phase'' in Fig.~\ref{strip_phase}).
For example, in the isotropic case $J_0=J_1=J$
the Dirac points are located at $\bm{q}=(0,\pm\pi/2)$.
As we approach the phase boundaries $J_1-J_0=\pm 2J$,
the two Dirac points come closer to each other and eventually 
merge at $\bm{q}=(0,0)$ for $J_1=J_0-2J$
and at $\bm{q}=(\pi/2,\pi/2)$ for $J_1=J_0+2J$.
On the other hand,
in the gapped phase where $J_0-2J>J_1$ or $J_0+2J<J_1$,
there is an energy gap
between positive and negative energy bands
(``A-phase'').

The effective Hamiltonian around the Dirac points is a
Dirac Hamiltonian with the mass terms proportional to
the next-nearest-neighbor hopping $K$.
Note that $\Theta(\bm{q})$ vanishes
at $\bm{q}=(0,0)$ and $(\pi/2,\pi/2)$.
This means that the band gap is closed at $J_1=J_0\pm2J$
even in the presence of the next-nearest-neighbor hopping.
Thus the parameter space $J_0/J-J_1/J$ is
divided into
three regions by the phase boundaries $J_1=J_0\pm2J$
(the dashed lines in Fig.~\ref{strip_phase}).

\begin{figure}[t]
 \includegraphics[width=84mm]{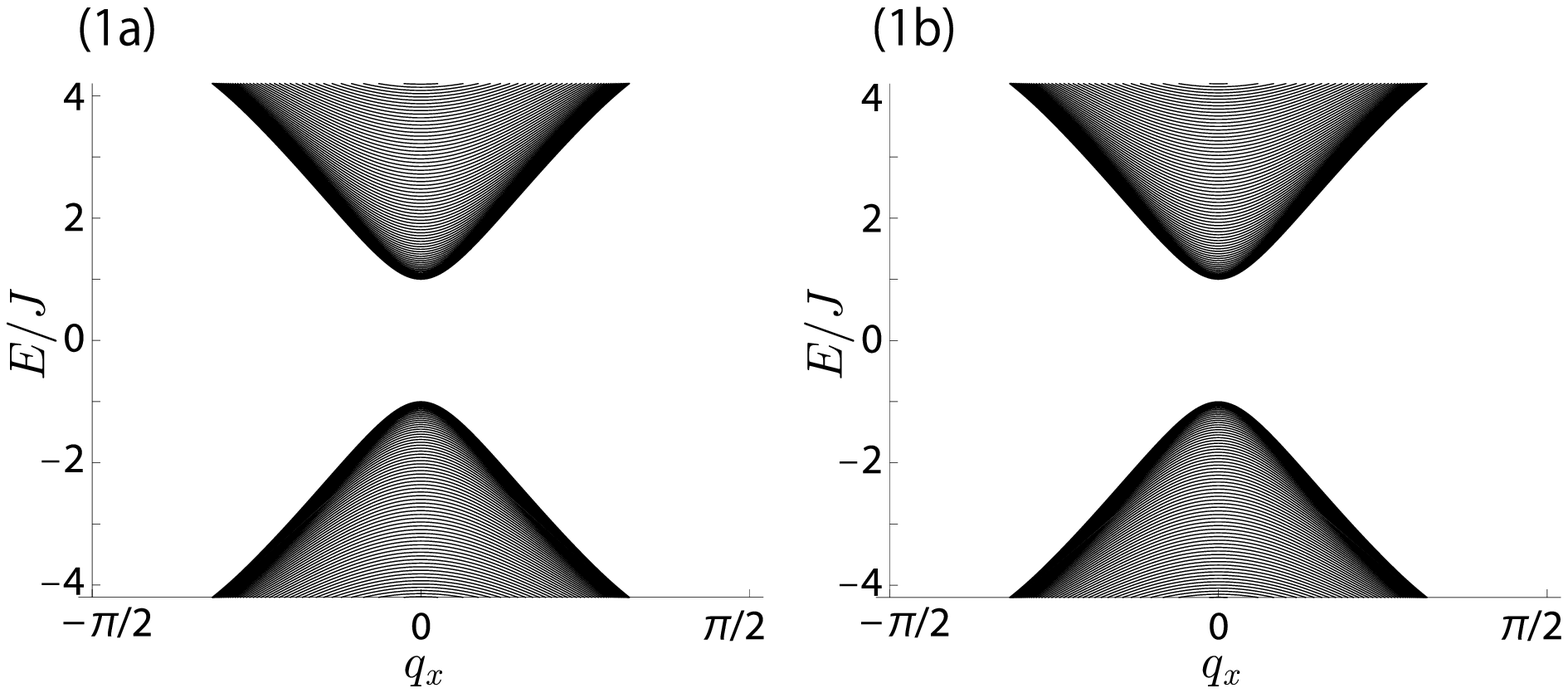}
 \vspace{0pt}
 \includegraphics[width=84mm]{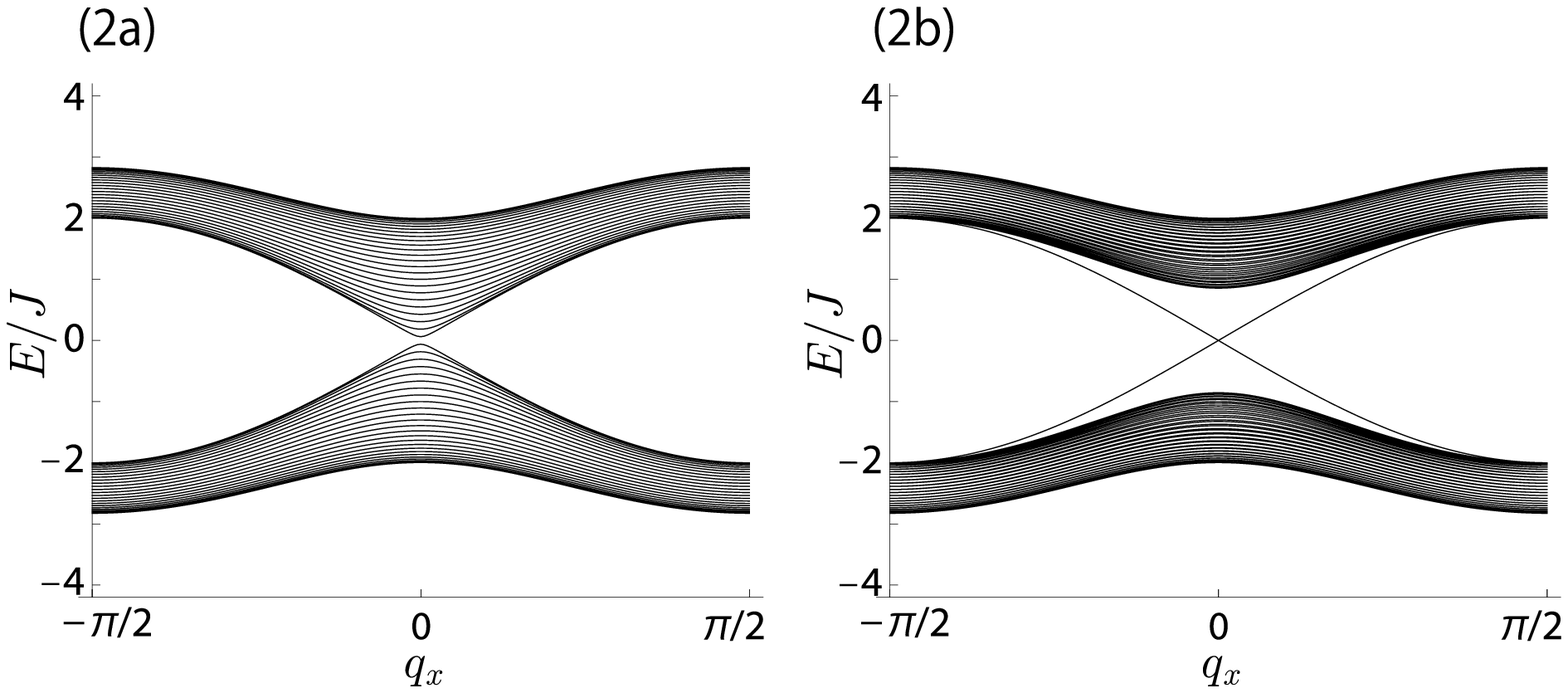}
 \vspace{0pt}
 \includegraphics[width=84mm]{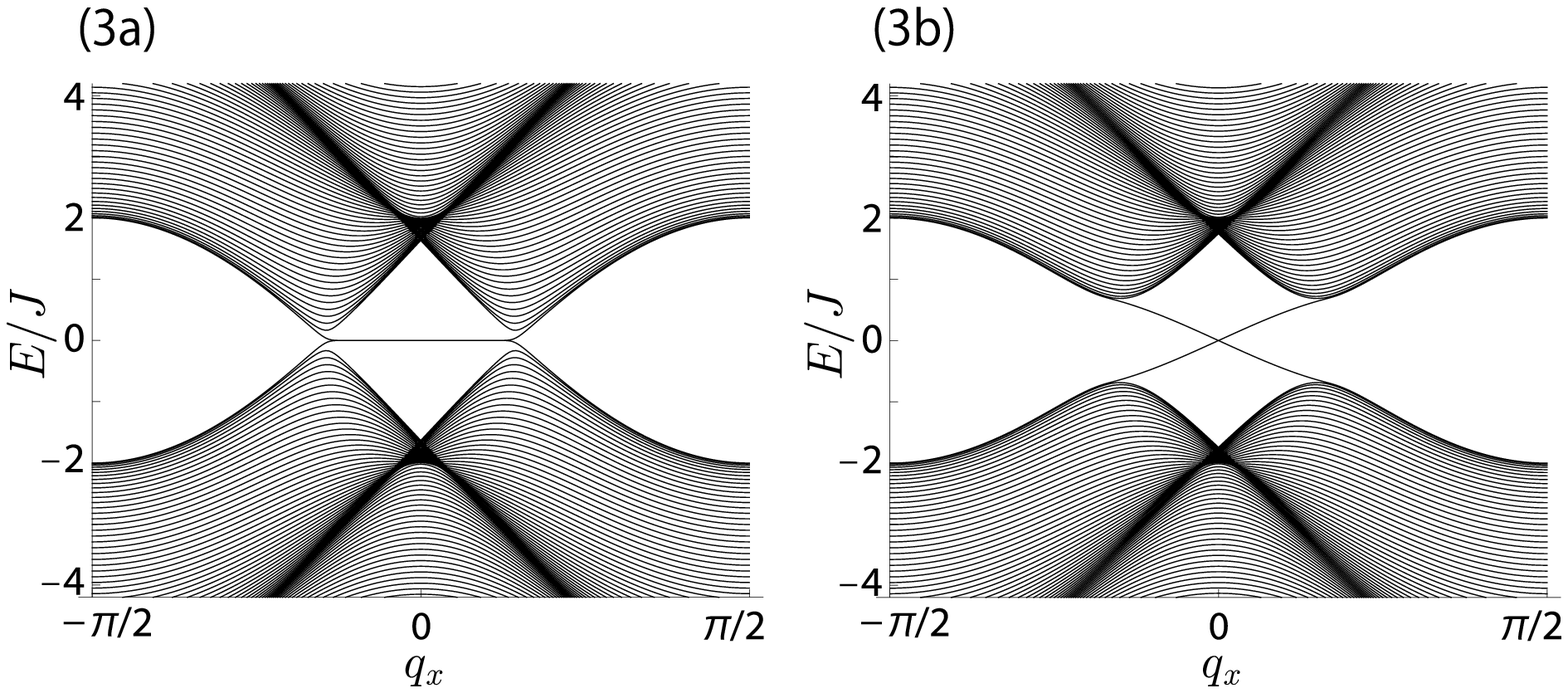}
 \vspace{0pt}
 \includegraphics[width=84mm]{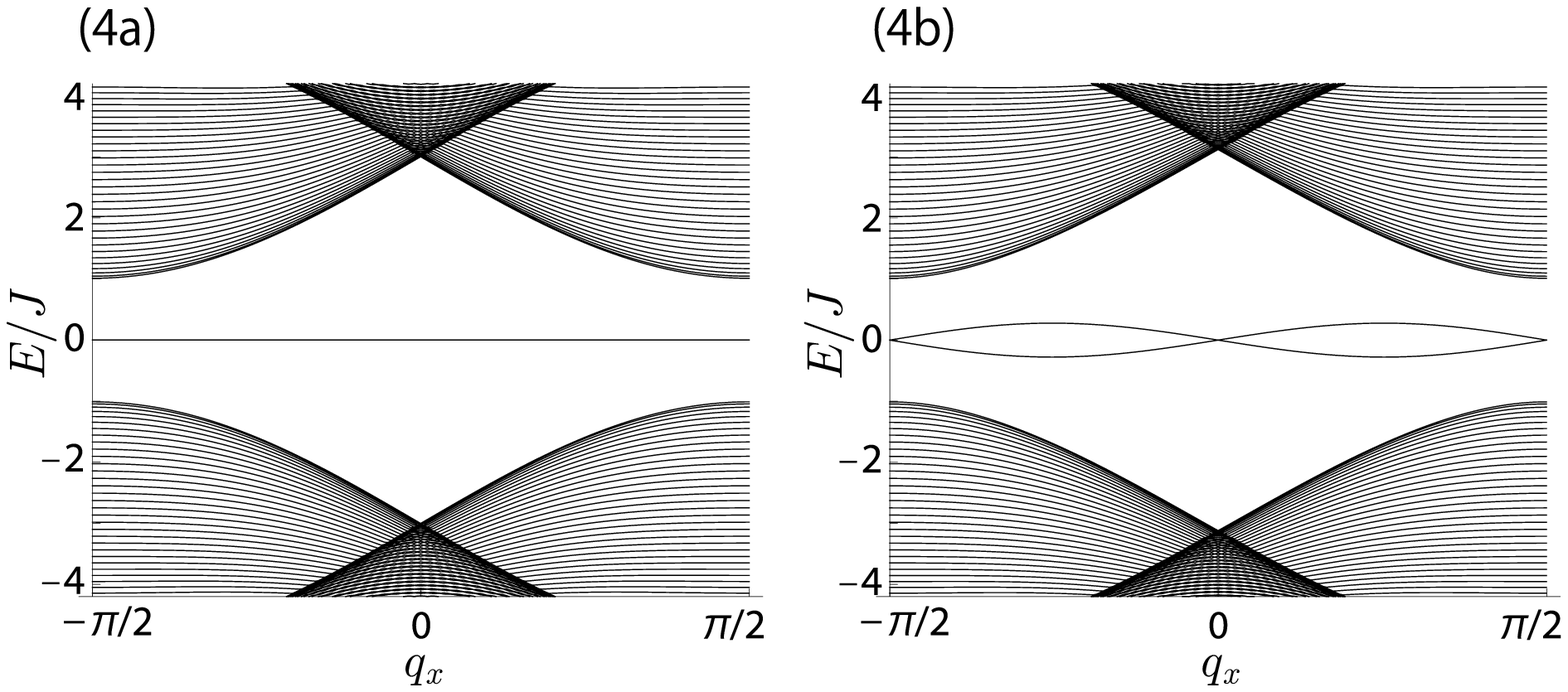}
 \vspace{-4pt}
 \caption{
Energy spectra of the one-dimensional strip
[Fig.\ \ref{strip_phase}(b)] as a function of the momentum along the edge.
The energy spectra are calculated for the following parameter sets
[see also Fig.~\ref{strip_phase}(a)]:
$J_{\mu}/J=(4,1,1,1)$ in (1a) and (1b);
$J_{\mu}/J=(1,1,1,1)$ in (2a) and (2b);
$J_{\mu}/J=(3,3,1,1)$ in (3a) and (3b);
$J_{\mu}/J=(1,4,1,1)$ in (4a) and (4b).
The $\mathbb{Z}_2$ gauge are fixed as $u^{\mu}=(-1,1,1,1)$.
The second-nearest-neighbor hopping $K^\alpha_i=0$ ($\alpha=1,2$ and $i=z,x$)
in the left figures [(1a), (2a), (3a), (4a)],
$K^\alpha_i/J=0.15$ 
in the right figures [(1b), (2b), (3b), (4b)].
}
 \label{energy_spectrum-2Dstrip}
\end{figure}

We have numerically diagonalized the Majorana tight-binding model
$\mathcal{H}_0+\mathcal{H}_z+\mathcal{H}_x$ for the strip geometry,
shown in Fig.~\ref{strip_phase}(b),
where the edges are parallel to the link
vectors $\bm{e}_0$ and $\bm{e}_2$.
The energy spectra of $\mathcal{H}_0$ in the strip geometry
are shown in Fig.~\ref{energy_spectrum-2Dstrip}
for 
$J_{\mu}/J=(4,1,1,1)$ [(1a) and (1b)],
$J_{\mu}/J=(1,1,1,1)$ [(2a) and (2b)],
$J_{\mu}/J=(3,3,1,1)$ [(3a) and (3b)],
and $J_{\mu}/J=(1,4,1,1)$ [(4a) and (4b)].
We have chosen two values for the next-nearest-neighbor hopping:
$K=0$ 
[(1a),(2a),(3a),(4a)] and 
$K/J=0.15$ 
[(1b),(2b),(3b),(4b)].

Without the next-nearest-neighbor hopping terms ($K=0$),
flat bands appear exactly at zero energy
in the region $J_0-2J<J_1$.
In the gapless phase (B-phase), the energy spectrum shows two Dirac points
at time-reversal symmetric momenta,
and the doubly degenerate zero-energy flat bands
connect these two points through $q_x=0$
(i.e., not through $q_x=\pi/2$); see Fig.~\ref{energy_spectrum-2Dstrip} (3a).
In the gapped phase (A-phase) where $J_0+2J<J_1$,
the bulk bands are fully gapped, and the flat bands
are extended in the whole Brillouin zone
[Fig.~\ref{energy_spectrum-2Dstrip} (4a)].

The existence of these zero-energy flat bands can be explained
by a topological argument.\cite{Ryu2002}
When $K=0$, $\lambda^4$ and $\lambda^5$ decouples in our model.
The bulk Hamiltonian for $\lambda^4$ 
(or $\lambda^5$) has the form
\begin{align}
 h(\bm{q})=\bm{R}(\bm{q})\cdot \bm{\sigma},
\end{align}
where $\bm{R}(\bm{q})$ is a two-dimensional vector,
\begin{align}
 \bm{R}(\bm{q})=
 \begin{pmatrix}
  -\text{Im}[\widetilde\Phi(\bm{q})] \\
  -\text{Re}[\widetilde\Phi(\bm{q})]
 \end{pmatrix}
\end{align}
with
\begin{align}
\widetilde\Phi(\bm{q})=
e^{-iq_x}(-J_0e^{iq_x}+J_1e^{iq_y}+J_2e^{-iq_x}+J_3e^{-iq_y}),
\label{tildePhi(q)}
\end{align}
$\bm{q}$ is the wave vector in the first Brillouin zone, and 
$\bm{\sigma}=(\sigma^x,\sigma^y)$.
In Eq.\ (\ref{tildePhi(q)}) $\widetilde\Phi(\bm{q})$ has
the phase factor $e^{-iq_x}$ [cf.~$\Phi(\bm{q})$ in Eq.\ (\ref{Phi(q)})],
because we have chosen the unit cell depicted by the dashed line
in Fig.~\ref{strip_phase}(b) which is commensurate with the
presence of the boundary.
Note that $h(\bm{q})$ has chiral symmetry, $\{h(\bm{q}),\sigma^z\}=0$.
When we fix $q_x$, $h(\bm{q})|_{q_x}$ can be regarded as a
one-dimensional Hamiltonian with wave number $q_y$
in the direction perpendicular to the edge.
The one-dimensional Hamiltonian has zero-energy edge modes if a loop
trajectory that $\bm{R}(\bm{q})|_{q_x}$ draws as $q_y$ is varied
encloses the origin $\bm{R}=0$ in the two-dimensional parameter
space $\bm{R}=(R^x,R^y)$.\cite{Ryu2002,note_zeromode}
The number of zero-energy edge modes is given by the winding number
of the loop and can change only when 
the loop touches the origin, i.e., when a band gap closes.
For a fixed $q_x$, the loop is an ellipse described by the equation
\begin{align}
 \frac{[R^x+(J_0+J_2)\sin q_x]^2}{(J_1-J_3)^2}+
 \frac{[R^y+(J_0-J_2)\cos q_x]^2}{(J_1+J_3)^2}=1.
\label{ellipse}
\end{align}
[In deriving Eq.~(\ref{ellipse}) we have ignored the
phase factor $e^{-iq_x}$ in $\widetilde\Phi(\bm{q})$,
since it does not change the winding number.]
The flat band appears when the origin $\bm{R}=0$ is inside the
ellipse, i.e., when $q_x$ satisfies
\begin{align}
 \sin^2 q_x <\frac{(J_1-J_3)^2[(J_1+J_3)^2-(J_0-J_2)^2]}
 {4(J_0J_1+J_2J_3)(J_0J_3+J_1J_2)}. \label{flat-band_condition}
\end{align}
In the phase diagram in Fig.~\ref{strip_phase}(a),
the right-hand side of Eq.\ (\ref{flat-band_condition}) is
larger than unity for $J_1>J_0+2J$ (A-phase in the upper side),
while it is less than zero for $J_1<J_0+2J$ (A-phase in the lower side),
which explains the spectra shown in
Fig.~\ref{energy_spectrum-2Dstrip} (1a) and (4a).
Otherwise, when $J_0-2J<J_1<J_0+2J$ (B-phase), the right-hand side of
(\ref{flat-band_condition}) takes an intermediate value between 0 and 1,
corresponding to the Fig.~\ref{energy_spectrum-2Dstrip} (2a) and (3a).

When the next-nearest-neighbor terms are included 
($K\neq 0$),
the bulk bands are gapped in the whole region 
of the A- and B-phases.
Then the flat bands are 
split from the zero energy,
except at
the time-reversal invariant momenta $q_x=0,\pi/2$.
Hence, the edge modes in 
the B-phase have
a single zero-energy point in the
first Brillouin zone [Fig.~\ref{energy_spectrum-2Dstrip} (2b) and (3b)], 
since the flat bands for $K=0$
pass through $q_x=0$ only,
while those in the
A-phase have an even number of
zero-energy points
[Fig.~\ref{energy_spectrum-2Dstrip} (1b) and (4b)].

\subsection{$\mathbb{Z}_2$ index}

The phase with topologically protected edge states
is characterized by a nontrivial $\mathbb{Z}_2$ index calculated in the bulk.
The $\mathbb{Z}_2$ index introduced by Kane and Mele\cite{KaneMele,FuKane}
for time-reversal invariant band insulators in class AII is defined
through the matrix
\begin{align}
w_{ij}(\bm{q})=\langle u_i(-\bm{q})|\mathcal{T}u_j(\bm{q})\rangle,
\end{align}
where $|u_j(\bm{q})\rangle$ is the single-particle Bloch
wave function in the $i$-th filled bands.
The $\mathbb{Z}_2$ invariant $\nu$ is then given by
\begin{align}
\nu=
 \prod_{\bm{q}:\text{TRIM}}
 \frac{\sqrt{\text{det}[w(\bm{q})]}}{\text{Pf}\,[w(\bm{q})]},
\label{Z_2 index}
\end{align}
where TRIM is the time-reversal invariant momenta in the Brillouin
zone, $(0,0)$, $(\pi/2,\pi/2)$, $(0,\pi)$, and $(-\pi/2,\pi/2)$.  The
sign of the square root in the numerator is chosen to be continuous
along the path connecting the four time-reversal invariant momenta.
The topological phase of our model in class DIII can also be
characterized by the $\mathbb{Z}_2$ index defined by
Eq.\ (\ref{Z_2 index}).

For models in symmetry class DIII,
the $\mathbb{Z}_2$ index becomes apparent when the Hamiltonian is
expressed in the off-diagonal form by utilizing the chiral symmetry.
Let us introduce the operator $Q$ that has the eigenvalue $+1$ $(-1)$
for the states in the empty (filled) band of the BdG Hamiltonian
$\tilde{\chi}$.
When $\tilde{\chi}(\bm{q})$ is diagonalized as 
$\tilde{\chi}=
V\text{diag}(\epsilon_1,\epsilon_2,\cdots,-\epsilon_1,-\epsilon_2,\cdots)
V^{-1}$
with a unitary matrix $V$,
the operator $Q$ is given by 
\begin{align}
 Q=V\text{diag}(1,1,\cdots,-1,-1,\cdots)V^{-1}.
\end{align}
In the basis where $\tilde{\chi}(\bm{q})$
takes the off-diagonal form of Eq.\ (\ref{chiral_BdG}),
the operator $Q$ also takes the form
\begin{align}
 Q(\bm{q})=
 \begin{pmatrix}
  0 & q(\bm{q}) \\
  q^{\dagger}(\bm{q}) & 0
 \end{pmatrix}.
\label{projection}
\end{align}
The off-diagonal component $q(\bm{q})$
satisfies the relations $q^T(-\bm{q})=-q(\bm{q})$ [so does $D(\bm{q})$]
and $q^{\dagger}q=qq^{\dagger}=I$ (from $Q^2=I$),
where $I$ is a unit matrix.
In this basis the operator $Q$ is related to the BdG Hamiltonian
$\tilde{\chi}$ by
\begin{equation}
Q(\bm{q})=\frac{1}{\varepsilon_{\bm{q}}}\tilde{\chi}_D(\bm{q})
\label{Q and chi_D}
\end{equation}
with $\varepsilon_{\bm{q}}$ defined in Eq.\ (\ref{varepsilon_q}).

The eigenvectors of $Q$ in Eq.\ (\ref{projection})
are given by
\begin{align}
u_{a\pm}(\bm{q})=
\frac{1}{\sqrt2}
 \begin{pmatrix}
  n_a \\ \pm q^{\dagger}(\bm{q})n_a
 \end{pmatrix}
 \qquad (a=1,2),
 \label{ev-projection}
\end{align}
where $\pm$ indicates the eigenvalue $\pm1$
(i.e., empty and filled bands), and $n_a$ are unit vectors,
\begin{equation}
n_1=\begin{pmatrix} 1 \\ 0 \end{pmatrix},
\qquad
n_2=\begin{pmatrix} 0 \\ 1 \end{pmatrix}.
\end{equation}
Since the eigenspace of the operator $Q$ of the eigenvalue $-1$ is
the same as the Hilbert space that spanned by the filled bands of
the BdG Hamiltonian, the $\mathbb{Z}_2$ index calculated with
the vectors in Eq.\ (\ref{ev-projection}) is equal to that calculated
for the original BdG Hamiltonian.

Applying the time-reversal operator $\mathcal{T}=it^y\mathcal{K}$
to the eigenvector of the filled states given in Eq.\ (\ref{ev-projection}) yields
\begin{equation}
\mathcal{T}|u_{a-}(\bm{q})\rangle
=\frac{1}{\sqrt2}
\begin{pmatrix}
-q^T(\bm{q})n_a \\ -n_a
\end{pmatrix}.
\end{equation}
The matrix $w$ is then obtained as
\begin{align}
 w_{ab}(\bm{q})=\langle u_{a-}(-\bm{q})|\mathcal{T}u_{b-}(\bm{q})\rangle
=-q_{ba}(\bm{q}).
\end{align}
It then follows from Eqs.\ (\ref{chiral_BdG}), (\ref{D(q)}), and
(\ref{Q and chi_D}) that
\begin{equation}
w(\bm{q})=\frac{1}{\varepsilon_{\bm{q}}}
\begin{pmatrix}
\Theta(\bm{q}) & -i\Phi^*(\bm{q}) \\
i\Phi(\bm{q}) & -\Theta^*(\bm{q})
\end{pmatrix}.
\end{equation}
Note that $\mathrm{det}[w(\bm{q})]=-1$ in the whole Brillouin zone,
and that $w(\bm{q})$ becomes a purely imaginary antisymmetric matrix
at the TRIM.
Hence the $\mathbb{Z}_2$ index in Eq.\ (\ref{Z_2 index}) is reduced to
\begin{equation}
\nu=\prod_{\bm{q}:\mathrm{TRIM}}\mathrm{sgn}[\Phi(\bm{q})].
\end{equation}
At the TRIM we have
\begin{subequations}
\begin{align}
&\Phi(0,0)=-J_0+J_1+J_2+J_3,\\
&\Phi(0,\pi)=J_0+J_1-J_2+J_3,\\
&\Phi(\pi/2,\pi/2)=J_0-J_1+J_2+J_3,\\
&\Phi(-\pi/2,\pi/2)=-J_0-J_1-J_2+J_3.
\end{align}
\label{Phi at TRIM}
\end{subequations}
The isotropic point, $J_\mu=J$, has $\nu=-1$ and is thus a topologically
nontrivial state.
In the limits where one of $J_\mu$ is much larger than the other three,
$\nu=+1$ and the ground state is topologically trivial.
These results are in agreement with the numerical results presented
in Sec.~\ref{sec:energy spectrum}.
The presence or absence of helical Majorana edge states is
dictated by the $\mathbb{Z}_2$ index $\nu$.

At the phase boundaries between the topologically nontrivial phase
and trivial phases, $\Phi(\bm{q})$ vanishes at least 
at one of the TRIM.
Using Eqs.\ (\ref{Phi at TRIM}), we arrive at the phase diagram
shown in Fig.~\ref{phase-diagram}, in which the shaded regions are
topologically trivial phases and the rest is a topological phase
(except on the phase boundaries).

\section{Edge states and spin correlation function}
\label{sec:spin_correlation}

In this section we examine spin correlations of the ground state of
the model, especially spin correlation functions along the edge of
the two-dimensional system.
The edge states appear in time-reversal pairs
and form the helical Majorana edge modes as in the time reversal helical
$p$-wave superconductors.\cite{Schnyder2008}
Spin correlation functions are calculated for these helical edge states.

Before proceeding to the calculation of correlation functions,
we examine which operators 
have non-vanishing expectation values 
in the ground state.
Operators can vanish due to two reasons:
the projection operator and the constants of motion. 
The former one restricts non-vanishing operators to be the product of
an even number of the Majorana operators on each site,
because
\begin{align}
 \frac{1+D_j}{2}\lambda_{j}^{p}
 \frac{1+D_j}{2}=\frac{1+D_j}{2}\frac{1-D_j}{2}\lambda_{j}^{p}=0,
\end{align}
where 
$p=0,\ldots,5$.
The latter one implies that non-vanishing operators can contain
the Majorana operators $\lambda^{\mu}$ with $\mu=0,1,2,3$
only in the form that does not flip the $\mathbb{Z}_2$ flux $\{\tilde{W}_p\}$,
since such Majorana operators alter the $\mathbb{Z}_2$ gauge configuration:
\begin{align}
 &u_{jk}^\mu\lambda_j^{\mu}=-\lambda_j^{\mu}u_{jk}^\mu \quad(\mu=0,1,2,3),\\
 &u_{jk}\lambda_j^{4}=\lambda_j^{4}u_{jk},\qquad
  u_{jk}\lambda_j^{5}=\lambda_j^{5}u_{jk},
\end{align}
and the projection operator does not flip the flux.
These conditions can be restated as follows:
non-vanishing operators for the ground state $|\text{GS}\rangle$
of the $\gamma$ matrix
Hamiltonian are
\begin{enumerate}
 \item $\mathbb{Z}_2$ gauge operators on closed strings of links,
 \item $\mathbb{Z}_2$ gauge operators on open strings of links, each string
       having either $\lambda_j^4$ or $\lambda_j^5$ at the both ends,
 \item ${i}\lambda_j^4\lambda_j^5$,
 \item $D_j$,
\end{enumerate}
and products thereof.
The product of the $\mathbb{Z}_2$ gauge field operators on closed strings
is rewritten as the
product of the plaquette operators $W_p$ and gives extra minus sign, 
since the plaquette operators are
the integrals of motion for both the Majorana Hamiltonian
and the $\gamma$ matrix Hamiltonian.
The correlation of operators which do not satisfy the above conditions 
does not extend beyond the nearest neighbor.

Here we consider the two-point correlation function of a local spin operator,
which is a single-site operator,
i.e., an operator that does not have a ``string'' composed
of a product of operators.
The only non-trivial
single-site Hermitian operator that satisfies the above conditions
is $i\lambda_j^4\lambda_j^5$.
The term $i\lambda_j^4\lambda_j^5$ is equal to
$i\gamma^5_j\gamma^0_j=
-i\alpha_j^{\mu}\zeta_j^{\mu}=
(\sigma^0\otimes\tau^2)_j$ in the $\gamma$
matrix and Pauli matrix representation, and is  
also written
as $2c_j^{\dagger}c_j-1$ 
in terms of
the complex fermions
(\ref{complex_fermion}).
In the bulk, 
the two-point correlation function of $i\lambda_j^4\lambda_j^5$
is short-ranged since $\lambda_j^4$ and $\lambda_j^5$
are free Majorana fermion operators and the bulk is gapped.
On the other hand,
the correlation function of $i\lambda^4\lambda^5$
along the edge is 
expected to decay algebraically 
in the topological phase.

Let us consider 
the semi-infinite system with 
the edge along the $x$ axis, 
and let $q_x$ be the momentum along the edge, which is conserved. 
The fermion operators can be expanded as
\begin{align}
 \begin{pmatrix}
a^4_{\boldsymbol{r}} \\
b^4_{\boldsymbol{r}} \\
a^5_{\boldsymbol{r}} \\ 
b^5_{\boldsymbol{r}} 
 \end{pmatrix}
& =
\int^{\pi/2}_{-\pi/2} 
dq_x
\sum_i
e^{{i} q_x r_{x}}
f_{q_x,i}(r_{y})
\xi_{q_x,i},
\nonumber \\
& =
\int^{\pi/2}_{0} 
dq_x
\sum_i
\left[
e^{{i} q_x r_{x}}
f_{q_x,i}(r_{y})
\xi_{q_x,i}
\right.
\nonumber \\
&\qquad
\qquad
\qquad
\qquad
\left.  
+
e^{-{i} q_x r_{x}}
f^*_{q_x,i}(r_y)
\xi^{\dag}_{q_x,i}
\right],
\end{align}
where 
$\boldsymbol{r}=(r_{x},r_{y})$ labels unit cells
which contain two sites
(the semi-infinite system is defined for $r_y<0$),
$f_{q_x,i}(r_y)$
is the $i$-th exact single particle 
wavefunction 
with momentum $q_x$ and 
energy $E_{q_x,i}$,
and
$f^*_{q_x,i}(r_y)$ is the particle-hole conjugate
with 
$-E_{q_x,i}$;
$\xi^{\ }_{q_x,i}$ ($\xi^{\dag}_{q_x,i}$)
is the fermion annihilation (creation)
operator associated with these levels.
In computing the correlation function on the edge, 
the dominant contributions come only from the
modes localized at the edge. 
Namely, the fermion operators near the edge 
are approximated by, at low energies,
\begin{align}
&
 \begin{pmatrix}
a^4_{\boldsymbol{r}} \\
b^4_{\boldsymbol{r}} \\
a^5_{\boldsymbol{r}} \\ 
b^5_{\boldsymbol{r}} 
 \end{pmatrix}
\simeq 
\int^{\pi/2}_0 
dq_x
\left[
e^{{i} q_x r_{x}}
g_{+,q_x}(r_{y})
\gamma_{+,q_x}
+
\mathrm{h.c.}
\right]
\nonumber \\
&\quad
\quad 
\quad 
\quad 
+
\int^{\pi/2}_0 
dq_x
\left[
e^{{i} q_x r_{x}}
g_{-,q_x}(r_{y})
\gamma_{-,q_x}
+
\mathrm{h.c.}
\right],
\end{align}
where 
$g_{\pm, q_x}(r_y)$
is 
the single-particle 
wavefunction of
the left (right)-moving edge mode
with momentum 
$q_x$,
and
$\gamma_{\pm, q_x}$
is the corresponding fermion annihilation operator.
The edge contribution to the Hamiltonian is given by
\begin{align}
 \mathcal{H}_{\text{edge}}
 &=
\int^{\pi/2}_0 
dq_x
E(q_x)\left(
\gamma_{+,q_x}^{\dagger}\gamma_{+,q_x}
-\gamma_{-,q_x}^{\dagger}\gamma_{-,q_x}
\right).
\end{align}
The energy dispersion
$E(q_x)$ for the edge mode is linear 
around 
a TRIM, $q_{x}=0$, 
as shown by the numerics
in Fig.~\ref{energy_spectrum-2Dstrip}. 

At $q_x=0$, the edge states are doubly degenerate at $E=0$.
For $K_z^1=K_z^2=K_z/2$ and $K_x^1=K_x^2=K_x/2, $
the zero-energy eigen wavefunctions
can be explicitly written as
\begin{align}
 g_{q_x=0}^{\alpha}(r_y)
 =(\Lambda_1^{r_y}-\Lambda_2^{r_y})\psi_0^{\alpha}
 \qquad
 (\alpha=1,2), \label{edge_wf_0}
\end{align}
where $r_y<0$ is an integer,
\begin{align}
 &\psi_0^{1}=
 \begin{pmatrix}
  -K_z \\
  (J_1u^1-J_3u^3)/2\pm A\\
  -K_x \\
  0
 \end{pmatrix}, \\
 &\psi_0^{2}=
 \begin{pmatrix}
  -K_x\\
  0\\
  K_z \\
  (J_1u^1-J_3u^3)/2\pm A
 \end{pmatrix},
\end{align}
with $A=\sqrt{K_z^2+K_x^2+(J_1u^1-J_3u^3)^2/4}$,
and $\Lambda_{1,2}$
are solutions of 
\begin{align}
 &\left(\frac{J_1u^1+J_3u^3}{2}\pm A\right)\Lambda^2
 +\left(J_0u^0e^{iq}+J_2u^2e^{-iq}\right)\Lambda
 \notag\\
 &\qquad+\left(\frac{J_1u^1+J_3u^3}{2}\mp A\right)=0.
\end{align}
When $|\Lambda_{1}|>1$ and $|\Lambda_{2}|>1$,
the wavefunctions
(\ref{edge_wf_0}) are normalizable and localized near the edges.
Such solutions of 
$\Lambda$ exist
when 
\begin{align}
 \left|\frac{J_1u^1+J_3u^3}{2}\pm A\right|<
 \left|\frac{J_1u^1+J_3u^3}{2}\mp A\right|
\label{edge-condition1}
\end{align}
and 
\begin{align}
 (J_0u^0+J_2u^2)^2>(J_1u^1+J_3u^3)^2, \label{edge-condition2}
\end{align}
where all $J_{\mu}$ are assumed to be positive.
The former condition (\ref{edge-condition1}) determines 
which signs to be taken. 
The latter condition (\ref{edge-condition2}) coincides with the 
region
in Fig.\ \ref{strip_phase} where there are 
zero-energy
edge states at $q_x=0$. 
In lowest order in $q_x$, 
the two-fold degeneracy of the zero modes
is lifted;
the energy dispersion near $q_x=0$ is 
$E=\pm vq$,
with the velocity 
\begin{align}
v=
\sqrt{(J_0u^2-J_2u^2)^2 (K_z^2+K_x^2)/A^2}. 
\label{edge_disp}
\end{align}
In lowest order in $q_x$, 
the eigen 
wavefunctions $g_{\pm,q_x}^{(0)}$
are linear combinations
of $g_{q_x=0}^1$ and $g_{q_x=0}^2$.

From the edge theory with a linear dispersion at low energies, 
one can immediately see
the (equal-time) two-point correlation function of the 
Majorana fermion operators decay along the edge as
$
\langle
\lambda^s(r_x)
\lambda^{s'}(r_x')
\rangle 
\sim
(r_x-r_x')^{-1}
$. 
The two-point correlation function of 
the operator 
$i\gamma^5\gamma^0=$
$\sigma^0 \otimes \tau^2 = i\lambda_j^4\lambda_j^5$ can be represented,
using the
Wick's theorem, as
\begin{align}
\langle (\sigma^0\otimes\tau^2)_r\,(\sigma^0\otimes\tau^2)_{r'}\rangle
&=
\langle (i\lambda^4\lambda^5)_r\,(i\lambda^4\lambda^5)_{r'}\rangle\notag\\
&\sim \frac{C(r_y,r_y')}{(r_x-r_x')^2},
\end{align}
where $C(r_y,r_y')$ is a function of  $r_y$ and $r_y'$ which is
determined by the wavefunction of the edge modes $g_{\pm,q_x}^{(0)}$
and decays exponentially into the bulk.

\section{Vortex bound states}
\label{sec:vortex}

In this section we discuss vortex bound states in the topological phase.
An isolated vortex in a
topological superconductor can accommodate
a topologically protected zero-energy Majorana state.\cite{Read00}
The time-reversal symmetry of our model implies that there are
two such Majorana zero-energy states 
which form a Kramers' doublet.

In our model a vortex corresponds to a plaquette with
a $0$-flux in the $\pi$-flux background.
Such $0$-flux excitations always appear in pair since the total flux is a
conserved quantity modulo $2\pi$.
As we noted above, each vortex should have two Majorana bound states.
To confirm the number of bound Majorana states, we have numerically
diagonalized Hamiltonian in the topological phase
(the B phase in Fig.~\ref{strip_phase}) for the system
size of $20\times 42$ sites, in which two $0$-flux plaquettes are
placed along the $y$ direction.
We have imposed periodic boundary conditions in the $x$ and $y$ directions
and set the parameters as $J^{\mu}=1$ and $K^{1,2}_{x,z}=0.3$.

\begin{figure}[t]
   \includegraphics[width=85mm]{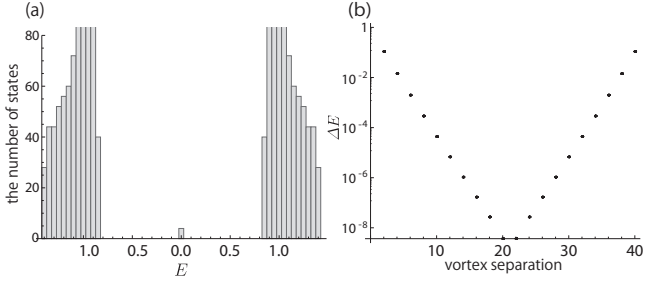}  
\caption{
(a) The number of states in the topological phase
for $20\times 42$ system with a pair of vortices
separated by 20 lattice spacings.
Periodic boundary conditions are imposed in the $x$ and $y$ directions.
The parameters in the Hamiltonian are $J^{\mu}=1$ and $K^{1,2}_{x,z}=0.3$.
(b) The energy difference $\varDelta E$ between the
positive and negative energy eigenvalues of bound states,
as a function of the distance between the vortices.}
 \label{vortex_bs}
\end{figure}

Figure \ref{vortex_bs}(a) shows the number of eigenstates
when two vortices are separated by 20 lattice spacings.
We find four nearly-zero-energy states inside the bulk gap,
i.e., two states per vortex.
The energy eigenvalues of these midgap states are
$\pm\varDelta E/2$,
each energy eigenvalue being two-fold degenerate.
Figure \ref{vortex_bs}(b) shows $\varDelta E$
as a function of the distance $r$ between the two vortices.
The dependence of $\varDelta E$ on $r$ is symmetric about $r=21$,
because of the periodic boundary conditions imposed.
The clear exponential dependence on $r$ ($r<21$) confirms that
the energy difference $\varDelta E$ is due to a small overlap
of exponential tails of wave functions bound to the two vortices.

\section{Extended Kitaev model on the one-dimensional lattice}
\label{sec:1D_model}

\begin{figure}[t]
 \includegraphics[width=20mm]{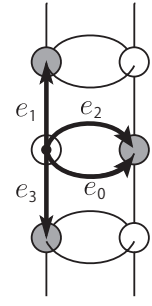}
 \caption{One-dimensional system on a two-leg cylindrical lattice.}
 \label{lattice-1D}
\end{figure}

In this section we study the extended Kitaev model on the cylinder geometry,
i.e.,
the ladder with two sets of rungs. 
The sites on the ladder
are divided into 
A and B sublattices,
shown as open and filled circles in 
Fig.~\ref{lattice-1D}, respectively.
The $\mu$-links are defined as in the two-dimensional case.

\begin{figure}[t]
 \includegraphics[width=60mm]{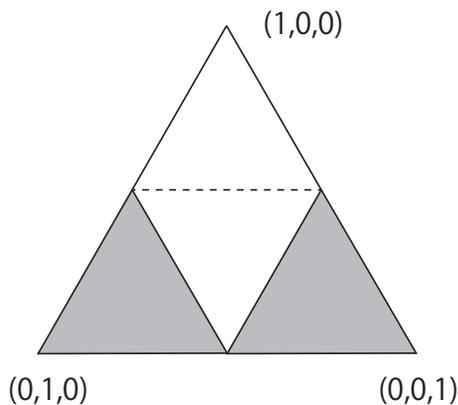}
 \caption{
Phase diagram of the one-dimensional cylindrical lattice model
in the parameter space $(J_0+J_2,J_1,J_3)$.
The phase in which one of the horizontal bonds $J_0$ and $J_2$ 
is greater
 than the other three bonds is a topologically trivial phase. 
The shaded regions (tetrahedra) 
are a topological phase, 
where each end of the ladder has two-fold degenerate zero-energy
Majorana states.
}
 \label{phase_1Dcyl}
\end{figure}

We consider the nearest-neighbor interaction Hamiltonian
\begin{align}
 \mathcal{H}_{\text{1D}}&=-\sum_{\mu=0}^3J_{\mu}
 \sum_{\mu\text{-links}}
 (\alpha_j^{\mu}\alpha_k^{\mu}+\zeta_j^{\mu}\zeta_k^{\mu}).
\end{align}
We use the Majorana fermion representation of the Dirac matrices
 (\ref{gamma-Majorana}), and 
combine $\lambda^4$ and $\lambda^5$ to make complex fermions as 
in Eq.\ (\ref{complex_fermion}),
and take the Fourier transform of the Majorana operators 
\begin{subequations}
 \begin{align}
 &A_q=\frac{1}{\sqrt{L}}\sum_{j\in\text{A}}e^{-iqy_j}
  (\lambda_j^4+i\lambda_j^5)/2 ,\\
 &B_q=\frac{1}{\sqrt{L}}\sum_{k\in\text{B}}e^{-iqy_k}
   (\lambda_k^4+i\lambda_k^5)/2,
\end{align}
\end{subequations}
where $L$ is the length of the cylinder,
 $y_j$ and $y_k$ are the positions of the sites in the vertical coordinate.
The Hamiltonian in the momentum space is then given by
\begin{align}
 \mathcal{H}_{\text{1D}}&=i\sum_{\mu=0}^3J_{\mu}\sum_{\mu\text{-links}}
 u_{jk}^{\mu}(\lambda_j^4\lambda_k^4+\lambda_j^5\lambda_k^5) \notag\\
 &=\sum_q \Psi_q^{\dagger}\chi_{\text{1D}}\Psi_q,
\end{align}
where the spinor $\Psi_q$ is
\begin{align}
 \Psi_q=
 \begin{pmatrix}
  A_q \\ B_q \\ A_{-q}^{\dagger} \\ B_{-q}^{\dagger}
 \end{pmatrix},
\end{align}
and 
\begin{align}
 \chi_{\text{1D}}=
 \begin{pmatrix}
  0 & i\Phi(q) & 0 & 0 \\
  -i\Phi^{\ast}(q) & 0 & 0 & 0 \\
  0 & 0 & 0 & i\Phi(q) \\
  0 & 0 & -i\Phi^{\ast}(q) & 0
 \end{pmatrix} \label{chi_1D}
\end{align}
with 
\begin{align}
 \Phi(q)=J_0u^0+J_1u^1e^{iq}+J_2u^2+J_3u^3e^{-iq}.
\end{align}
The eigenenergies of $\chi_{\text{1D}}$ are $E=\pm|\Phi(q)|$ and
each energy level is doubly degenerate. 
From the Lieb's theorem,
the ground state is obtained for the $\mathbb{Z}_2$ gauge field
configurations with 
$\pi$-flux per each square,
i.e., when the condition
$
\mathrm{sgn}\,
[
(J_1 u^1)(J_3 u^3)
(J_0 u^0 + J_2 u^2)
(J_0 u^0 + J_2 u^2)
]
=-1
$
is satisfied.
Without loss of generality, we will work with the $\mathbb{Z}_2$ gauge
$u^{\mu}=(1,-1,1,1)$. 
The ground-state energy is then a function of three parameters
$J_0+J_2$, $J_1$, and $J_3$.
Even without the next-nearest neighbor interaction terms,
the ground state is gapped except at the phase boundaries,
\begin{align}
 J_1-J_3=\pm(J_0+J_2).
 \label{1Dcyl_boundary}
\end{align}
The phase diagram is depicted in Fig.\ \ref{phase_1Dcyl}.

We diagonalized numerically 
the Hamiltonian 
for a finite length system 
with open boundary condition in the leg direction. 
No midgap states are found
if $J_1-J_3<J_0+J_2$ and $J_1-J_3>-(J_0+J_2)$ 
(non-shaded region in Fig.\ \ref{phase_1Dcyl}),
while 
midgap states bound to each end are found
when
$J_1-J_3>J_0+J_2$ and $J_1-J_3<-(J_0+J_2)$ 
(shaded regions in Fig.\ \ref{phase_1Dcyl}).
These midgap modes of Majorana fermions have two-fold degeneracy 
due to the time-reversal symmetry,
or equivalently, spin 1/2 degrees of freedom bound on each edge.

The Hamiltonian of one-dimensional system (\ref{chi_1D}) 
has the same symmetry as that of two-dimensional system (\ref{chi_tilde}).
Thus the $\mathbb{Z}_2$ invariant is the product of 
\begin{align}
 \frac{\sqrt{\text{det}[w(q)]}}{\text{Pf}[w(q)]}=\text{sgn}[\Phi(q)]
\end{align}
at the TRIM in the one-dimensional Brillouin zone,
\begin{subequations}
 \begin{align}
  &\Phi(0)=J_0-J_1+J_2+J_3,  \\
  &\Phi(\pi)=J_0+J_1+J_2-J_3.
\end{align}
\end{subequations}
The phase boundaries obtained from the $\mathbb{Z}_2$ invariant coincides
with those from numerics, which are already given in (\ref{1Dcyl_boundary}).
Since the class DIII Hamiltonian (\ref{chi_1D})
can be decomposed into two independent blocks,
each of which is a member of class AIII, 
the $\mathbb{Z}_2$ invariant in this case 
coincides with the even-odd parity of the 
integral invariant (winding number)
of class AIII for the blocks. 
In turn, the winding number can be obtained 
by drawing the loop trajectory 
in the parameter space
defined by $\Phi(q)$ in Eq.\ (\ref{chi_1D})
[see for discussion around Eq.\ (\ref{tildePhi(q)})];
The product 
$\text{sgn}[\Phi(0)]\text{sgn}[\Phi(\pi)]$
(i.e., the $\mathbb{Z}_2$ invariant)
then tells us that, when negative, the loop trajectory drawn by 
$\Phi(q)$ encloses the origin an odd number of times.

\section{Conclusions}

In this paper, we have introduced a time-reversal symmetric
two-dimensional quantum spin model in a topologically non-trivial gapped phase,
as a $\gamma$-matrix extension of the Kitaev model
on the square lattice.
Through a fermion representation of the $\gamma$ matrices,
this model is equivalent to a time-reversal symmetric 
two-dimensional topological superconductor 
(i.e., a system in class DIII in the Altland-Zirnbauer classification).
The Hamiltonian consists of nearest-neighbor interaction terms and
next-nearest-neighbor interaction terms, all of which are transformed to
free Majorana fermion hopping Hamiltonian with $\mathbb{Z}_2$ gauge field.
In the parameter space of the Hamiltonian, 
topologically trivial ground states and non-trivial ones are realized.
We have confirmed that these two phases are distinguished by
the $\mathbb{Z}_2$ topological invariant.

We have shown using both numerical and analytical methods
the existence of topologically protected, a Kramers' pair of
Majorana edge modes,
which is a hallmark of a time-reversal symmetric topological superconductor.
A local operator of a product of two flavors of Majorana operators,
which is equivalent to the density operator of a complex fermion,
has a nonvanishing correlation that decays in inverse-square of the distance
along the edge and decays exponentially in the bulk.
We have also shown numerically that a vortex of the $\mathbb{Z}_2$ gauge field
hosts a Kramers' pair of zero-energy Majorana states.

On the one-dimensional ladder lattice, we have constructed 
the same type of extended Kitaev model.
Similarly to the square lattice case, 
two topologically distinct types of ground states appear
in the phase diagram, which are characterized by
the $\mathbb{Z}_2$ topological invariant.

\acknowledgments
This work was supported in part by
Grant-in-Aid for JSPS Fellows (No.~227763) and
Grant-in-Aid for Scientific Research (No.~21540332)
from the Japan Society for the Promotion of Science
and by the National Science Foundation under Grant No.~NSF PHY05-51164.
AF and SR are grateful to the Kavli Institute for Theoretical Physics
for its hospitality, where this paper was completed.

\appendix* 

\section{Jordan-Wigner transformation of the gamma matrix Kitaev model}
\label{sec:Jordan_Wigner}

In this appendix, 
we present a solution to the 
Kitaev-type model (\ref{Hamiltonian-spin}),
in terms of the 
Jordan-Wigner transformation,
following the 
solution of the original Kitaev model
by Jordan-Wigner transformation. 
\cite{Feng2007,Yao2007,HanDongChen_JiangpingHu07,HanDongChen_Nussinov07}

\subsection{Jordan-Wigner transformation of the Dirac matrices}

As a first step,
we note that it is possible to represent 
the Dirac matrices for a given site 
in terms of two complex fermions $c_1$ and $c_2$
as
\begin{align}
 \begin{array}{l}
  \alpha^0=(c_1+c_1^{\dagger}),\\[+3pt]
  \alpha^2=(c_1-c_1^{\dagger})/i,
 \end{array}\quad
 \begin{array}{l}
  \alpha^1=(c_2+c_2^{\dagger}),\\[+3pt]
  \alpha^3=(c_2-c_2^{\dagger})/i,
 \end{array}
\label{alpha-Majorana}
\end{align}
where 
$\{c_1, c^{\dag}_1\}=\{c_2, c^{\dag}_2\}=1$,
and 
$\{c_1, c_2\}= \{c_1, c^{\dag}_2\}=0$. 
The right-hand side of four equations in (\ref{alpha-Majorana}) are Majorana
fermion operators.
Similarly, using $\zeta^{\mu}=\gamma^5\gamma^0\alpha^{\mu}$,
\begin{align}
 &\begin{array}{l}
  \zeta^0=e^{i\pi n_2}(c_1-c_1^{\dagger})/i,\\
  \zeta^2=-e^{i\pi n_2}(c_1+c_1^{\dagger}),
 \end{array}\quad
 \begin{array}{l}
  \zeta^1=e^{i\pi n_1}(c_2-c_2^{\dagger})/i,\\
  \zeta^3=-e^{i\pi n_1}(c_2+c_2^{\dagger}), 
 \end{array}
 \label{zeta-Majorana} \\
 &i\gamma^5\gamma^0=\sigma^0\otimes\tau^y=e^{i\pi(n_1+n_2)} ,\label{zeta-5}
\end{align}
where $n_{1}=c^{\dag}_1 c_1$, $n_{2}=c^{\dag}_2 c_2$,
and we have used the relation
$e^{i\pi n_a}=1-2n_a=-(c_a+c_a^{\dagger})(c_a-c_a^{\dagger})$.

\begin{figure}[t]
 \includegraphics[width=80mm]{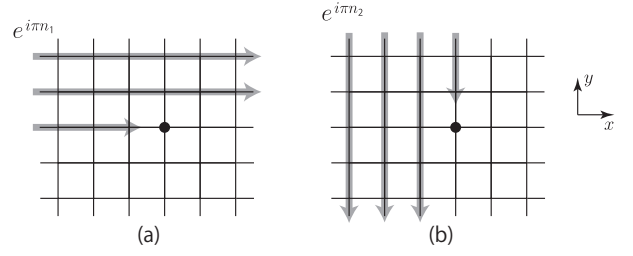}
\caption{String operators of Jordan-Wigner fermions that represent Dirac
 matrices for (a) Majorana fermion made of $c_1$ ($U_j$), 
 and (b) those made of $c_2$ ($V_j$).}
 \label{JW-fig}
\end{figure}

For the Kitaev-type model (\ref{Hamiltonian-spin}),
we need to prepare a set of two complex fermions 
$c_{1j}$ and $c_{2j}$ for each site labeled by $j$. 
Accordingly, 
one needs to introduce string operators 
to ensure commutation relations for operators 
sitting on different sites. 
At fist, we consider the string operators for 
zeroth and second components of the Dirac matrices,
which contain single Majorana operators made of $c_1$.
Since 0-links and 2-links horizontally connect neighboring sites
(Fig.~\ref{square_lattice}),
we define an order of sites on the two-dimensional lattice that runs
horizontally:
\begin{align}
 &(x_1,y_1)<(x_2,y_2)\,\Leftrightarrow\,\notag\\
 &\quad y_1>y_2 
 \text{ or }(y_1=y_2\text{ and }x_1<x_2), \label{JW-order1}
\end{align}
where $x$ and $y$ are the two-dimensional coordinates
[Fig.~\ref{JW-fig}(a)].  
Multiplying the right-hand side of the zeroth and second components
in Eqs.\ (\ref{alpha-Majorana}) and 
(\ref{zeta-Majorana}) by string operators of products of $e^{i\pi n_1}$ 
with the order (\ref{JW-order1}) [Fig.~\ref{JW-fig}(a)],
\begin{align}
 U_j&=\prod_{j>k}e^{i\pi n_{1k}}, \label{JW-string1}
\end{align}
makes these operators commute between different sites.

Next, we can also introduce string operators for the first and third components,
with another order of sites on the same two-dimensional
lattice [Fig.~\ref{JW-fig}(b)]
\begin{align}
 &(x_1,y_1)<(x_2,y_2)\,\Leftrightarrow\,\notag\\
 &\quad x_1<x_2 
 \text{ or }(x_1=x_2\text{ and }y_1>y_2), \label{JW-order2}
\end{align}
as
\begin{align}
 V_j=\prod_{j>k}e^{i\pi n_{2k}}. \label{JW-string2}
\end{align}
The latter string operator (\ref{JW-string2}) ensures commutativity
of the first and third
components of Dirac matrix on different sites.
Finally, in order for Jordan-Wigner fermion representation of 
all four components of Dirac matrix to be bosonic,
the first and third components of Dirac matrix on a site $j$ is
multiplied by
string operator that runs all the sites besides $j$:
\begin{align}
 W_j=\prod_{k(\neq j)}e^{i\pi n_{1k}}.
\end{align}
As a result, we obtain the Jordan-Wigner fermion representation
of Dirac matrices as
\begin{align}
 \begin{array}{l}
  \alpha_j^0=(c_{1j}+c_{1j}^{\dagger}) U_j,\\[+2pt]
  \alpha_j^1=(c_{2j}+c_{2j}^{\dagger}) V_jW_j,\\[+2pt]
  \alpha_j^2=-i(c_{1j}-c_{1j}^{\dagger}) U_j,\\[+2pt]
  \alpha_j^3=-i(c_{2j}-c_{2j}^{\dagger}) V_jW_j,
 \end{array}
\end{align}
and
\begin{align}
 \begin{array}{l}
  \zeta_j^0=-i(c_{1j}-c_{1j}^{\dagger})e^{i\pi n_{2j}} U_j,\\[+2pt]
  \zeta_j^1=-i(c_{2j}-c_{2j}^{\dagger})e^{i\pi n_{1j}} V_jW_j,\\[+2pt]
  \zeta_j^2=-(c_{1j}+c_{1j}^{\dagger})e^{i\pi n_{2j}} U_j,\\[+2pt]
  \zeta_j^3=-(c_{2j}+c_{2j}^{\dagger})e^{i\pi n_{1j}}V_jW_j.
 \end{array}
\end{align}

\subsection{Jordan-Wigner transformation of the $\gamma$ matrix Kitaev model}

By utilizing the Jordan-Wigner fermion representation of Dirac matrices,
we can transform our spin model to a free Majorana fermion model,
without redundancy.

On 0-links, since $j\in \text{A}$ and $k\in \text{B}$ connected by a 0-link 
have the order $j<k$ defined in (\ref{JW-order1}), 
the nearest-neighbor interaction terms are transformed as follows.
\begin{align*}
 \alpha_j^0\alpha_k^0
&=(c_{1j}+c_{1j}^{\dagger})U_j(c_{1k}+c_{1k}^{\dagger})U_k \notag\\
&=-(c_{1j}-c_{1j}^{\dagger}) (c_{1k}+c_{1k}^{\dagger}),\\
 \zeta_j^0\zeta_k^0
&=-(c_{1j}-c_{1j}^{\dagger})e^{i\pi n_{2j}}U_j
 (c_{1k}-c_{1k}^{\dagger})e^{i\pi n_{2k}}U_k \notag\\
&=\left[(c_{1j}+c_{1j}^{\dagger})e^{i\pi n_{2j}}\right]
 \left[(c_{1k}-c_{1k}^{\dagger})e^{i\pi n_{2k}}\right].
\end{align*}

Considering the order of $j\in \text{A}$ and $k\in \text{B}$ 
in the string operators ((\ref{JW-order1}) and (\ref{JW-order2})), 
nearest-neighbor interaction terms on the other three links are similarly
transformed as
\begin{align*}
 \alpha_j^1\alpha_k^1
 &=\left[(c_{2j}+c_{2j}^{\dagger})e^{i\pi n_{1j}}\right]
   \left[(c_{2k}-c_{2k}^{\dagger})/i\,e^{i\pi n_{1k}}\right], \\
 \zeta_j^1\zeta_k^1
 &=-(c_{2j}-c_{2j}^{\dagger})(c_{2k}+c_{2k}^{\dagger}),
\end{align*}
\begin{align*}
 \alpha_j^2\alpha_k^2
 &=-(c_{1j}-c_{1j}^{\dagger})(c_{1k}+c_{1k}^{\dagger}),\\
 \zeta_j^2\zeta_k^2
 &=\left[(c_{1j}+c_{1j}^{\dagger})e^{i\pi n_{2j}}\right]
 \left[(c_{1k}-c_{1k}^{\dagger})e^{i\pi n_{2k}}\right],
\end{align*}
\begin{align*}
 \alpha_j^3\alpha_k^3
 &=\left[(c_{2j}+c_{2j}^{\dagger})e^{i\pi n_{1j}}\right]
 \left[(c_{2k}-c_{2k}^{\dagger})e^{i\pi n_{1k}}\right], \\
 \zeta_j^3\zeta_k^3 
 &=-(c_{2j}-c_{2j}^{\dagger})(c_{2k}+c_{2k}^{\dagger}).
\end{align*}

Here we introduce two Majorana fermion operators on each site as
\begin{align}
 &\left\{
 \begin{array}{ll}
  \lambda_j^4&=-(c_{2j}+c_{2j}^{\dagger})e^{i\pi n_{1j}}, \\[+2pt]
  \lambda_j^5&=-i(c_{2j}-c_{2j}^{\dagger}),
 \end{array}
 \right. \\
 &\left\{
 \begin{array}{ll}
  \lambda_k^4&=-i(c_{2k}-c_{2k}^{\dagger})e^{i\pi n_{1k}},\\[+2pt]
  \lambda_k^5&=c_{2k}+c_{2k}^{\dagger},
 \end{array}
 \right.  \label{JW-lambda}
\end{align}
where $j\in\text{A}$ and $k\in\text{B}$.
Those Majorana fermion operators transform the nearest-neighbor
interaction terms on 
$1\text{-link}$s and $3\text{-link}$s to free Majorana hopping terms
without $\mathbb{Z}_2$ gauge operators, as
\begin{align}
& -\sum_{\mu=1,3}J_{\mu}\sum_{\mu\text{-links}}
(\alpha_j^{\mu}\alpha_k^{\mu}+\zeta_j^{\mu}\zeta_k^{\mu})
\nonumber \\
&\quad =
 i\sum_{\mu=1,3}J_{\mu}\sum_{\mu\text{-links}}
 (\lambda_j^4\lambda_k^4+\lambda_j^5\lambda_k^5).
\end{align}
Majorana fermions in Eq.\ (\ref{JW-lambda}) also transform the nearest-neighbor
interaction terms on $0\text{-link}$s and $2\text{-link}$s
to  free Majorana hopping terms, however in this case, 
with $\mathbb{Z}_2$ gauge operators:
\begin{align}
 & -\sum_{\mu=0,2}J_{\mu}\sum_{\mu\text{-links}}
(\alpha_j^{\mu}\alpha_k^{\mu}+\zeta_j^{\mu}\zeta_k^{\mu})
\nonumber \\
&\quad =
 i\sum_{\mu=0,2}J_{\mu}\sum_{\mu\text{-links}}u_{jk}
 (\lambda_j^4\lambda_k^4+\lambda_j^5\lambda_k^5),
\end{align}
where $u_{jk}$ are  $\mathbb{Z}_2$ gauge operators of the form
\begin{align}
 u_{jk}=-(c_{1j}+c_{1j}^{\dagger})(c_{2j}+c_{2j}^{\dagger})
 (c_{1k}-c_{1k}^{\dagger})(c_{2k}-c_{2k}^{\dagger}).
\end{align}
The $\mathbb{Z}_2$ gauge operators commute with each other,
and also commute with 
$\lambda_j^4,\lambda_k^4,\lambda_j^5,\lambda_k^5$:
\begin{align}
 &[u_{jk},u_{lm}]=0, \\
 &[u_{jk},\lambda_l^4]=[u_{jk},\lambda_l^5]=0.
\end{align}

Consequently, we obtain the Jordan-Wigner transformation of 
the $\gamma$ matrix Kitaev model as
\begin{align}
 \mathcal{H}
&
=i\left(\sum_{0\text{-links}}J_0u_{jk}+\sum_{1\text{-links}}J_1+
 \sum_{2\text{-links}}J_2u_{jk}+\sum_{0\text{-links}}J_3\right)
\nonumber \\
&
\quad \times 
(\lambda_j^4\lambda_k^4+\lambda_j^5\lambda_k^5).
\end{align}
In this representation, the number of the $\mathbb{Z}_2$ gauge operators
reduces to the half of that in the local Majorana
representation (\ref{Hamiltonian-Majorana}),
and therefore there is no redundancy in this fermionic representation.
If we consider the periodic boundary condition,
string operators appear in the Hamiltonian in connecting both edges,
as discussed in the case of the Kitaev model.\cite{Yao2007}


\begin{thebibliography}{99}
\bibitem{Note_Yao}
The gapless (algebraic spin liquid) phase in a similar model to ours is discussed in
	Ref.~\onlinecite{Yao08}.
\bibitem{note_zeromode}
In this case $h(\bm{q})$ can be continuously deformed to
a simple one-dimensional chiral-symmetric Hamiltonian which obviously has
zero-energy end states.
\end{thebibliography}
\end{document}